\documentclass[5p,authoryear,uniquename=false]{elsarticle}
\usepackage{epstopdf}
\usepackage{subfigure,graphicx,tabularx}
\usepackage{float}
\usepackage{amsmath, amsfonts, amssymb,mathrsfs}
\usepackage{amsthm}
\usepackage{epsf}
\usepackage{amsfonts}
\usepackage{stmaryrd}
\usepackage{amssymb}
\usepackage{leftidx}
\usepackage{color}
\usepackage{mathtools}
\usepackage{placeins}
\usepackage{booktabs}
\usepackage{enumitem}
\usepackage{caption}
\usepackage{multirow}
\usepackage{stackengine}
\usepackage[utf8]{inputenc}
\usepackage[english]{babel}
\usepackage{color}
\usepackage{pifont}
\usepackage{fontawesome}
\usepackage{hyperref}
\newcolumntype{?}{!{\vrule width 1pt}}

\DeclareMathAlphabet{\mathsfit}{T1}{\sfdefault}{\mddefault}{\sldefault}
\SetMathAlphabet{\mathsfit}{bold}{T1}{\sfdefault}{\bfdefault}{\sldefault}

\theoremstyle{plain}
\newtheorem{theorem}{Theorem}

\newtheorem{remark}[theorem]{Remark} 
\def\bfu{{\bf u}}

\def\bfx{{\bf x}}

\def\bfE{{\bf E}}

\def\bfI{{\bf I}}

\def\bfN{{\bf N}}

\def\bfS{{\bf S}}

\def\bfX{{\bf X}}

\def\bfe{{\bf e}}

\def\e0{\varepsilon_0}

\def\s0{\sigma_0}

\def\sts{\sigma_{\texttt{ts}}}
\def\scs{\sigma_{\texttt{cs}}}
\def\shs{\sigma_{\texttt{hs}}}

\long\def\symbolfootnote[#1]#2{\begingroup%
\def\thefootnote{\fnsymbol{footnote}}\footnote[#1]{#2}\endgroup}

\makeatletter
\renewcommand\@biblabel[1]{}
\renewenvironment{thebibliography}[1]
     {\section*{\refname}%
      \@mkboth{\MakeUppercase\refname}{\MakeUppercase\refname}%
      \list{}%
           {\leftmargin0pt
            \@openbib@code
            \usecounter{enumiv}}%
      \sloppy
      \clubpenalty4000
      \@clubpenalty \clubpenalty
      \widowpenalty4000%
      \sfcode`\.\@m}
     {\def\@noitemerr
       {\@latex@warning{Empty `thebibliography' environment}}%
      \endlist}
\makeatother

\begin{document}
\begin{frontmatter}

\title{Breaking Four-Point and Three-Point Bending Tests\vspace{0.1cm}}

\vspace{-0.1cm}

\author{Subhrangsu Saha}
\ead{saha15@illinois.edu}

\author{Jeffery R. Roesler}
\ead{jroesler@illinois.edu}

\author{Oscar Lopez-Pamies}
\ead{pamies@illinois.edu}

\address{Department of Civil and Environmental Engineering, University of Illinois, Urbana--Champaign, IL 61801, USA}

\begin{abstract}

\vspace{-0.1cm}

Since their initial standardizations in the 1930s and 1950s, the so-called four-point and three-point bending tests on unnotched beams have been embraced by practitioners as two popular methods to indirectly measure the tensile strength of concrete, ceramics, and other materials with a large compressive strength relative to their tensile strength. This is because of the ease that the tests afford in both the preparation of the specimen (a beam of rectangular cross section) and the application of the loads (simple supports pressing on the specimen). Yet, this practical advantage has to be tempered by the fact that the observations from both of these tests --- being \emph{indirect} experiments in the sense that they involve \emph{not} uniform uniaxial tension but non-uniform triaxial stress states throughout the specimen --- have to be appropriately interpreted to be useful. By making use of the phase-field fracture theory initiated by \cite*{KFLP18}, which has been recently established as a complete theory of fracture capable of accurately describing the nucleation and propagation of cracks in elastic brittle materials under arbitrary quasistatic loading conditions, the main objective of this paper is to carry out a thorough 3D quantitative analysis of when and where fracture nucleates and propagates in four-point and three-point bending tests and thereby establish how to appropriately interpret their results. The focus is on the fundamental case of materials that can be considered homogeneous, isotropic, linear elastic brittle at the length scale of the beam. As a corollary, the analysis provides an explanation for why four-point bending tests typically yield smaller flexural strengths than three-point bending tests, a source of constant headaches for practitioners who have been left to wonder which test --- if any --- would be more appropriate for their purposes. The final section of this paper presents simple formulas for deducing the uniaxial tensile strength of a material directly from the flexural strength measured in each of these tests.

\endkeyword

\end{abstract}

\end{frontmatter}

\section{Introduction}

The use of bending tests on unnotched beams to study the strength of materials that exhibit weak tensile strength relative to their compressive strength --- such as for instance mortar, concrete, rocks, and ceramics --- dates back centuries; see, e.g., the classical work of \cite{Rankine1858}. In contrast to standard direct tests aimed at subjecting specimens to \emph{spatially uniform uniaxial tension} with the help of specialty grips, such tests are particularly accessible both in terms of specimen preparation and application of the loading. This practical advantage has to be tempered by the fact that the observations from bending tests --- being \emph{indirect} experiments in the sense that they involve \emph{not} uniform uniaxial tension but non-uniform triaxial stress states throughout the specimen --- have to be appropriately interpreted to be useful. 

The classical interpretation of fracture in bending tests --- which remains the one \emph{de facto} advocated across all pertinent ASTM standards to date --- is that fracture occurs whenever the maximum principal stress at any material point in the specimen reaches a critical value.\footnote{For concrete, this value is typically labeled by $R$ and referred to as the modulus of rupture or flexural strength. For other materials, such as ceramics, the label $S$ and terminology of flexural strength are standard; see, e.g., \citep{ASTMC78,ASTMC293,ASTMC1161}. In this work, for reasons that will become apparent below, we shall denote the flexural strength by $S_{max}$.} Such a maximum principal stress is presumed to be given by the Euler-Bernoulli formula and hence to be attained at the boundary of the specimen at any point(s) along the span where the maximum moment happens to be reached. This interpretation is nothing more than a direct application of Lam\'e's pioneering postulate for fracture nucleation \citep{Lame1831}, a postulate that has been long debunked. Indeed, over the past many decades, a multitude of experiments have repeatedly shown that fracture in a body where the stress field is \emph{not} spatially uniform does \emph{not} nucleate when the stress at a single material point violates a critical value. Instead, fracture nucleates once a stress violation has taken place over a sufficiently large region  of the body, the size and shape of this region being dependent on the geometry of the body, the material that is made of, and the specifics of the applied loading conditions. 

The classical interpretation of fracture in bending tests is thus incorrect. This fundamental misunderstanding has led to persistent confusion among practitioners and researchers alike, perhaps most clearly illustrated by the common, yet perplexing, belief that four-point bending tests yield smaller ``tensile strength'' than three-point bending tests; see, e.g., \citep{Wright1952,Lindner1955,Walker1957}.

\begin{figure}[t!]
\centering
\includegraphics[width=0.75\linewidth]{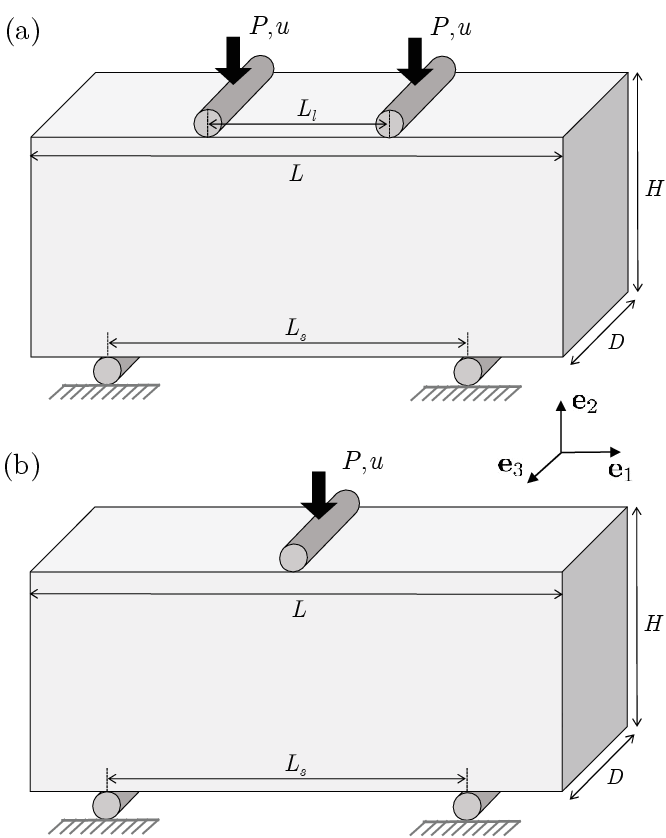}
\caption{{\small Schematics of the (a) four-point bending and (b) three-point bending fracture tests. Except for Section \ref{Sec: Final Comments}, where additional sizes are considered, the beam dimensions used throughout this study are fixed at $L=1$ m, $D=H=0.2$ m, $L_s=4 H=0.8$ m, and $L_l=L_s/2=0.4$ m.}}\label{Fig1}
\end{figure}

In this context, the objective of this paper is to bring resolution to the interpretation of the experimental results from bending tests by providing a thorough 3D quantitative description and explanation of when and where fracture nucleates and propagates in the specimens. The focus is on the two most prominent types of bending tests, namely, the so-called four-point and three-point bending tests depicted schematically in Fig.~\ref{Fig1}, this for the fundamental case of materials that can be considered homogeneous, isotropic, linear elastic brittle at the length scale of the specimens, within the setting of quasistatic loading conditions. We do so by deploying the phase-field fracture theory initiated by \cite*{KFLP18}, which has been recently established as a complete theory of fracture capable of accurately describing the nucleation and propagation of cracks in elastic brittle materials under arbitrary quasistatic loading conditions. In particular, we make use of the specialization of this theory to isotropic linear elastic brittle materials presented in \citep{KBFLP20,LDLP24}.
%
%

In a nutshell, the theory of \cite*{KFLP18} corresponds to a generalization of the classical phase-field regularization \citep{Bourdin00} of the variational theory of brittle fracture of \cite*{Francfort98}, which in turn corresponds to the mathematical statement of Griffith's fracture postulate in its general form of energy cost-benefit analysis \citep{Griffith21}. Consistent with the vast experimental evidence that has been amassed for over a century on numerous ceramics, metals, and polymers alike, the generalization consists in accounting for the strength of the material at large. Critically, the definition of strength --- a macroscopic material property that had long been mistreated and misunderstood (see, e.g., Sections 5.1.2 and 2.1 in \cite{BCLLP24} and \cite{LPDFL25}) --- is as follows: the strength of an elastic brittle material is the set of all critical stresses $\boldsymbol{\sigma}$ at which the material fractures when it is subjected to a state of monotonically increasing, spatially uniform, but otherwise arbitrary stress. Such a set of critical stresses defines a surface 
\begin{equation}\label{F-Strength}
\mathcal{F}(\boldsymbol{\sigma})=0
\end{equation}
in stress space, which is referred to as \emph{the strength surface of the material}. Precisely, the theory dictates that:
\begin{enumerate}[label=\textbf{\roman*.}]

\item{cracks nucleate and propagate only in regions where the strength surface (\ref{F-Strength}) of the material has been exceeded; and}

\item{they do so in a way that minimizes the sum of the potential and the surface energies.}
\end{enumerate}
In other words, the violation of the strength surface (\ref{F-Strength}) is a necessary but \emph{not} sufficient condition for crack nucleation and propagation. Sufficiency is established through the minimization of the sum of the potential and surface energies. 

A string of recent works \citep{KRLP18,KLP20,KBFLP20,KLP21,KRLP22,KLDLP24,KKLP24,KLP25,LPK25,KZDLP26} have provided a wide range of validation results for different materials (silicone, titania, graphite, polyurethane, PMMA, alumina, natural rubber, glass), specimen geometries (with large and small pre-existing cracks, V notches, U notches, and smooth boundaries), and loading conditions demonstrating that the phase-field fracture theory initiated by \cite*{KFLP18} indeed provides a complete framework for the description of fracture nucleation and propagation in nominally elastic brittle materials under arbitrary quasistatic loads.

The paper is organized as follows. We begin in Section \ref{Sec: Strength} by presenting full-field elastostatics results for when and where the strength surface (\ref{F-Strength}) is violated in the four-point and three-point bending tests as a function of the applied loading. \emph{Inter alia}, this elementary strength analysis is aimed at illustrating the differences between the stress fields that develop in both tests prior to the nucleation of fracture. In Section \ref{Sec: Fracture}, we spell out the governing equations that describe the nucleation and propagation of fracture in the bending tests, according to the phase-field theory. In Section \ref{Sec: Results}, we solve these equations numerically and present a series of representative results that illustrate when, where, and why beams fracture under four-point and three-point bending.  We conclude this work by recording a summary of our findings and a number of final comments in Section \ref{Sec: Final Comments}. They include simple formulas for deducing the uniaxial tensile strength of a material directly from the flexural strength measured in four-point and three-point bending tests.

\section{An elementary strength analysis of the bending tests}\label{Sec: Strength}

\subsection{Initial configuration and kinematics}\label{Sec: Initial Conf}

Consider a prismatic beam of length $L=1$ m and square cross section of depth and height $D=H=0.2$ m in the $\bfe_1$ and $\bfe_2$-$\bfe_3$ directions; see Fig. \ref{Fig1}. These specific values for $L$, $D$, and $H$ are chosen here because they are sufficiently large\footnote{In general, the smallest dimension ($D$ and/or $H$) of the beam should be chosen to be at least 10 times larger than the characteristic length scale of the underlying microstructure, e.g., the size ($\sim5$ mm) of the aggregates in mortar, so that the beam can be considered to be made of a homogeneous material.} and consistent with the ASTM standards for concrete \citep{ASTMC293,ASTMC78}. The Cartesian basis $\{\bfe_i\}$ stands for the laboratory frame of reference. Its origin is placed at the bottom center of the beam, so that, in its initial (undeformed and stress-free) configuration at time $t=0$ the beam occupies the domain
\begin{equation*}
\Omega=\left\{\bfX:|X_1|< \dfrac{L}{2},\, 0<X_2< H,\,|X_3|< \dfrac{D}{2}\right\}.
\end{equation*}
Making use of standard notation, we denote the boundary of the beam by $\partial\Omega$ and its outward unit normal by $\bfN$. 

At a later time $t\in(0,T]$, in response to the applied boundary conditions described below in Subsection \ref{Sec: BCs}, the position vector $\bfX$ of a material point in the beam will move to a new position specified by
\begin{equation*}
\bfx = \bfX+\bfu(\bfX,t),
\end{equation*}
where $\bfu(\bfX,t)$ is the displacement field. We write the associated strain at $\bfX$ and $t$ as
\begin{equation*}
\bfE(\bfu):=\dfrac{1}{2}\left(\nabla\bfu+\nabla\bfu^T\right).
\end{equation*}
\subsection{Material constitutive behavior: Elasticity, strength, and critical energy release rate}

The beam is taken to be made of a homogeneous, isotropic, linear elastic brittle material. Its mechanical behavior is hence characterized by three intrinsic macroscopic properties: (\emph{i}) its elasticity; (\emph{ii}) its strength; and (\emph{iii}) its critical energy release rate.

\paragraph{Elasticity} Precisely, granted isotropy, the elastic behavior of the material is characterized by the stored-energy function
\begin{equation}
W(\bfE(\bfu)) =\dfrac{E}{2(1+\nu)}\bfE\cdot\bfE+\dfrac{E\nu}{2(1+\nu)(1-2\nu)}({\rm tr}\,\bfE)^2,\label{W-E-nu}
\end{equation}
where $E>0$ and $\nu\in[-1,1/2]$ are the Young's modulus and Poisson's ratio, or, equivalently, by the stress-strain relation
\begin{equation}
\boldsymbol{\sigma}(\bfX,t)=\dfrac{E}{1+\nu}\bfE+\dfrac{E\,\nu}{(1+\nu)(1-2\nu)}({\rm tr}\,\bfE)\bfI.\label{S-E-nu}
\end{equation}

\paragraph{Strength} For definiteness, the strength surface of the material is taken to be characterized by the Drucker-Prager strength surface
\begin{equation}
\mathcal{F}(\boldsymbol{\sigma})=\sqrt{J_2}+\dfrac{\sigma_{\texttt{cs}}-\sigma_{\texttt{ts}}}
{\sqrt{3}\left(\sigma_{\texttt{cs}}+\sigma_{\texttt{ts}}\right)} I_1-\dfrac{2\sigma_{\texttt{cs}}\sigma_{\texttt{ts}}}
{\sqrt{3}\left(\sigma_{\texttt{cs}}+\sigma_{\texttt{ts}}\right)}=0,\label{DP-1}
\end{equation}
where $I_1={\rm tr}\,\boldsymbol{\sigma}$ and $J_2=\dfrac{1}{2}{\rm tr}\,\boldsymbol{\sigma}^2_{D}$, with $\boldsymbol{\sigma}_{D}=\boldsymbol{\sigma}-\dfrac{1}{3}({\rm tr}\,\boldsymbol{\sigma})\bfI$, stand for two of the standard invariants of the stress tensor $\boldsymbol{\sigma}$, while the constants $\sigma_{\texttt{ts}}>0$ and $\sigma_{\texttt{cs}}>0$ denote the uniaxial tensile and compressive strengths of the material, that is, they denote the critical stress values at which fracture nucleates under uniform uniaxial tension $\boldsymbol{\sigma}={\rm diag}(\sigma>0,0,0)$ and uniaxial compression $\boldsymbol{\sigma}={\rm diag}(\sigma<0,0,0)$, respectively.

\begin{remark}
\emph{Note that, according to our choice of signs in (\ref{DP-1}), any stress state such that
\begin{equation*}
\mathcal{F}(\boldsymbol{\sigma})\geq 0 
\end{equation*}
is in violation of the strength of the material.
}
\end{remark}
\begin{remark}
\emph{The two-material-parameter strength surface (\ref{DP-1}) is arguably the simplest model that has proven capable of describing reasonably well the strength of many nominally elastic brittle materials, thus its use here as a representative template.}
\end{remark}

\begin{remark}
\emph{The strength of a material is inherently stochastic. This is because the strength at any given macroscopic material point depends on the varying nature of the underlying defects where fracture originates. Consequently, the strength constants $\sts$ and $\scs$ in (\ref{DP-1}) should be considered as stochastic material constants, and not as deterministic values. We will come back to this important point in Subsection \ref{SubS-Stochasticity} below.}
\end{remark}

\paragraph{Critical energy release rate} Finally, the material resistance to crack growth is characterized by the critical energy release rate --- also termed the intrinsic fracture energy or toughness --- and is denoted by the non-negative constant 
\begin{equation*}
G_c.
\end{equation*}
From a physical standpoint, $G_c$ quantifies the energy per unit fracture area expended in the creation of new surface.

In the analysis that follows, for definiteness, we will present results for various mortar mixtures with the material constants listed in Table \ref{Table1}.
\begin{table}[H]\centering\small
\caption{Primary material constants for the mortar mixtures used as prototypical materials in this work (additional material constants utilized in the analysis are provided in Subsections \ref{SubS-lc} and \ref{SubS-Stochasticity}  and in Section \ref{Sec: Final Comments}).}
\begin{tabular}{l?cc}
\toprule
Elasticity constants & $E$ (GPa)& $\nu$ \\
\midrule
                     & $27$   & $0.20$ \\
\midrule
\midrule
Strength constants   & $\sts$ (MPa) & $\scs$ (MPa) \\
\midrule
                     & $4$                  & $20,40,60$               \\
\midrule
\midrule
Critical energy release rate   & $G_c$ (N/m) &   \\
\midrule
                               & $5,15,30,60$ & \\
\bottomrule
\end{tabular} \label{Table1}
\end{table}

\subsection{Boundary conditions}\label{Sec: BCs}

Consistent again with the ASTM recommendations, we consider that the beam rests on two simple frictionless supports at a distance --- typically referred to as the span length of the beam --- $L_s=4H=0.8$ m from each other; see Fig.~\ref{Fig1}. We describe these supports by the mixed boundary condition 
\begin{equation}\label{BC-B}
\left\{\begin{array}{l}
\sigma_{1j}(\bfX,t)N_j=0\vspace{0.2cm}\\
u_2(\bfX,t)=0\vspace{0.2cm}\\
\sigma_{3j}(\bfX,t)N_j=0
\end{array}\right.,\quad (\bfX,t)\in\partial\Omega_{\mathcal{B}}\times[0,T],
\end{equation}
where $\partial\Omega_{\mathcal{B}}=\partial\Omega_{\mathcal{B}_1}\cup\partial\Omega_{\mathcal{B}_2}$ with
\begin{equation*}
\left\{\begin{array}{l}
\partial\Omega_{\mathcal{B}_1}=\left\{\bfX:\left|X_1+\dfrac{L_s}{2}\right|<l,\, X_2=0\, , |X_3|<\dfrac{D}{2}\right\}\vspace{0.2cm}\\
\partial\Omega_{\mathcal{B}_2}=\left\{\bfX:\left|X_1-\dfrac{L_s}{2}\right|<l,\, X_2=0\, , |X_3|<\dfrac{D}{2}\right\}
\end{array}\right.
\end{equation*}
and $l=2.5$ mm.

\paragraph{Four-point bending} For the four-point bending test, see Fig.~\ref{Fig1}(a), the beam is loaded by two pressing supports symmetrically placed at a distance $L_l=2 H=0.4$ m from each other, which we describe by the mixed boundary condition 
\begin{equation}\label{BC-T-4p}
\left\{\begin{array}{l}
\sigma_{1j}(\bfX,t)N_j=0\vspace{0.2cm}\\
u_2(\bfX,t)=-u\vspace{0.2cm}\\
\sigma_{3j}(\bfX,t)N_j=0
\end{array}\right.,\quad (\bfX,t)\in\partial\Omega^{4p}_{\mathcal{T}}\times[0,T],
\end{equation}
where $\partial\Omega^{4p}_{\mathcal{T}}=\partial\Omega_{\mathcal{T}_1}\cup\partial\Omega_{\mathcal{T}_2}$,
\begin{equation*}
\left\{\begin{array}{l}
\partial\Omega_{\mathcal{T}_1}=\left\{\bfX:\left|X_1+\dfrac{L_l}{2}\right|<l,\, X_2=H\, , |X_3|<\dfrac{D}{2}\right\}\vspace{0.2cm}\\
\partial\Omega_{\mathcal{T}_2}=\left\{\bfX:\left|X_1-\dfrac{L_l}{2}\right|<l,\, X_2=H\, , |X_3|<\dfrac{D}{2}\right\}
\end{array}\right. 
\end{equation*}
with $u$ prescribed. The remaining part of the boundary $\partial\Omega\setminus\partial\Omega_{\mathcal{B}}\cup\partial\Omega^{4p}_{\mathcal{T}}$ is traction free, that is,
\begin{equation}\label{BC-free-4p}
\boldsymbol{\sigma}(\bfX,t)\bfN=\textbf{0},\quad(\bfX,t)\in \partial\Omega\setminus\partial\Omega_{\mathcal{B}}\cup\partial\Omega^{4p}_{\mathcal{T}}\times[0,T].
\end{equation}
For later use, observe that in terms of the resultant force
\begin{equation*}
2P=-\displaystyle\int_{\partial\Omega_{\mathcal{T}_1}}\sigma_{2j}N_j\,{\rm d}\bfX-\displaystyle\int_{\partial\Omega_{\mathcal{T}_2}}\sigma_{2j}N_j\,{\rm d}\bfX,
\end{equation*}
the resultant moment between the pressing supports ($X_1\in[-L_l/2,L_l/2]$) reads
\begin{equation}\label{M4p}
M=\dfrac{P L_s}{4}.
\end{equation}

\paragraph{Three-point bending} For the three-point bending test, see Fig.~\ref{Fig1}(b), the beam is loaded by just one pressing support at the midspan, which we describe by the mixed boundary condition 
\begin{equation}\label{BC-T-3p}
\left\{\begin{array}{l}
\sigma_{1j}(\bfX,t)N_j=0\vspace{0.2cm}\\
u_2(\bfX,t)=-u\vspace{0.2cm}\\
\sigma_{3j}(\bfX,t)N_j=0
\end{array}\right.,\quad (\bfX,t)\in\partial\Omega^{3p}_{\mathcal{T}}\times[0,T],
\end{equation}
where
\begin{equation*}
\partial\Omega^{3p}_{\mathcal{T}}=\left\{\bfX:|X_1|<l,\, X_2=H\, , |X_3|<\dfrac{D}{2}\right\}
\end{equation*}
with, again, $u$ prescribed. The remaining part of the boundary $\partial\Omega\setminus\partial\Omega_{\mathcal{B}}\cup\partial\Omega^{3p}_{\mathcal{T}}$ is traction free, that is,
\begin{equation}\label{BC-free-3p}
\boldsymbol{\sigma}(\bfX,t)\bfN=\textbf{0},\quad(\bfX,t)\in \partial\Omega\setminus\partial\Omega_{\mathcal{B}}\cup\partial\Omega^{3p}_{\mathcal{T}}\times[0,T].
\end{equation}
In this case, in terms of the resultant force
\begin{equation*}
P=-\displaystyle\int_{\partial\Omega^{3p}_{\mathcal{T}}}\sigma_{2j}N_j\,{\rm d}\bfX,
\end{equation*}
note that the resultant moment at the center of the beam ($X_1=0$) is also given by 
\begin{equation}\label{M3p}
M=\dfrac{P L_s}{4}.
\end{equation}

\subsection{The governing equations of elastic deformation}

Prior to the nucleation of fracture, neglecting inertia and body forces, the combination of the balance of linear momentum 
\begin{equation*}
{\rm Div}[\boldsymbol{\sigma}(\bfX,t)]=\textbf{0},\quad (\bfX,t)\in\Omega\times[0,T],
\end{equation*}
with the constitutive relation (\ref{S-E-nu}), the boundary conditions (\ref{BC-B}), (\ref{BC-T-4p}), (\ref{BC-free-4p}) for the four-point bending test and (\ref{BC-B}), (\ref{BC-T-3p}), (\ref{BC-free-3p}) for the three-point bending test, and the initial condition 
$$\bfu(\bfX,0)=\textbf{0}$$
constitute the initial-boundary-value problems that describe the elastic response of the  beam under both bending tests. 

These are standard initial-boundary-value problems that can be readily solved via the finite element (FE) method. The FE results that follow make use in particular of linear tetrahedral elements of minimum size $\texttt{h}=1$ mm, which is sufficiently small to generate converged solutions for the displacement field $\bfu(\bfX,t)$, the strain field $\bfE(\bfu(\bfX,t))$, and the stress field $\boldsymbol{\sigma}(\bfX,t)$ in the beam.

\begin{remark}
\emph{Here, it is important to note that while a 2D idealization may be tempting to reduce computational cost, it is insufficient for the present study. This is because neither plane-stress nor plane-strain approximations can accurately describe the violation of the strength surface of the material. Consequently, the use of these simplifications would lead to inaccurate predictions for the nucleation of fracture.}
\end{remark}

\subsection{When and where the strength surface is violated}

As a first step to gain insight into the first instance of fracture nucleation in the bending tests, we now present results for the evolution of the \emph{the strength violation set}
\begin{equation}\label{VF}
\mathcal{V}_{\mathcal{F}}(t):=\left\{\bfX:\mathcal{F}(\boldsymbol{\sigma}(\bfX,t))\geq 0\right\},
\end{equation}
that is, the set of all material points at which the strength surface $\mathcal{F}(\boldsymbol{\sigma})=0$ of the material is exceeded at time $t$, where $\boldsymbol{\sigma}(\bfX,t)$ is determined from the pertinent FE solutions of elastostatics, as the loading increases.

Since the moments (\ref{M4p}) and (\ref{M3p}) for the four-point and three-point bending tests are given by the same expression, it proves helpful to present the results \emph{not} in terms of the time $t$, or the applied displacement $u$ at the pressing supports, but in terms of the global stress measure
\begin{equation*}
S:=\dfrac{6M}{DH^2}.
\end{equation*}
%
%
%

All the results that follow pertain to mortars with the material constants listed in Table \ref{Table1}. 

\begin{remark}
\emph{Recall that in elastic brittle materials cracks can only nucleate and propagate in regions where the strength surface $\mathcal{F}(\boldsymbol{\sigma})=0$ of the material has been violated and so the results for (\ref{VF}) presented here provide a preliminary indicator of when and where cracks may first nucleate.
}
\end{remark}

\subsubsection{Four-point vs. three-point bending}

%
\begin{figure}[t!]
\centering
\includegraphics[width=0.925\linewidth]{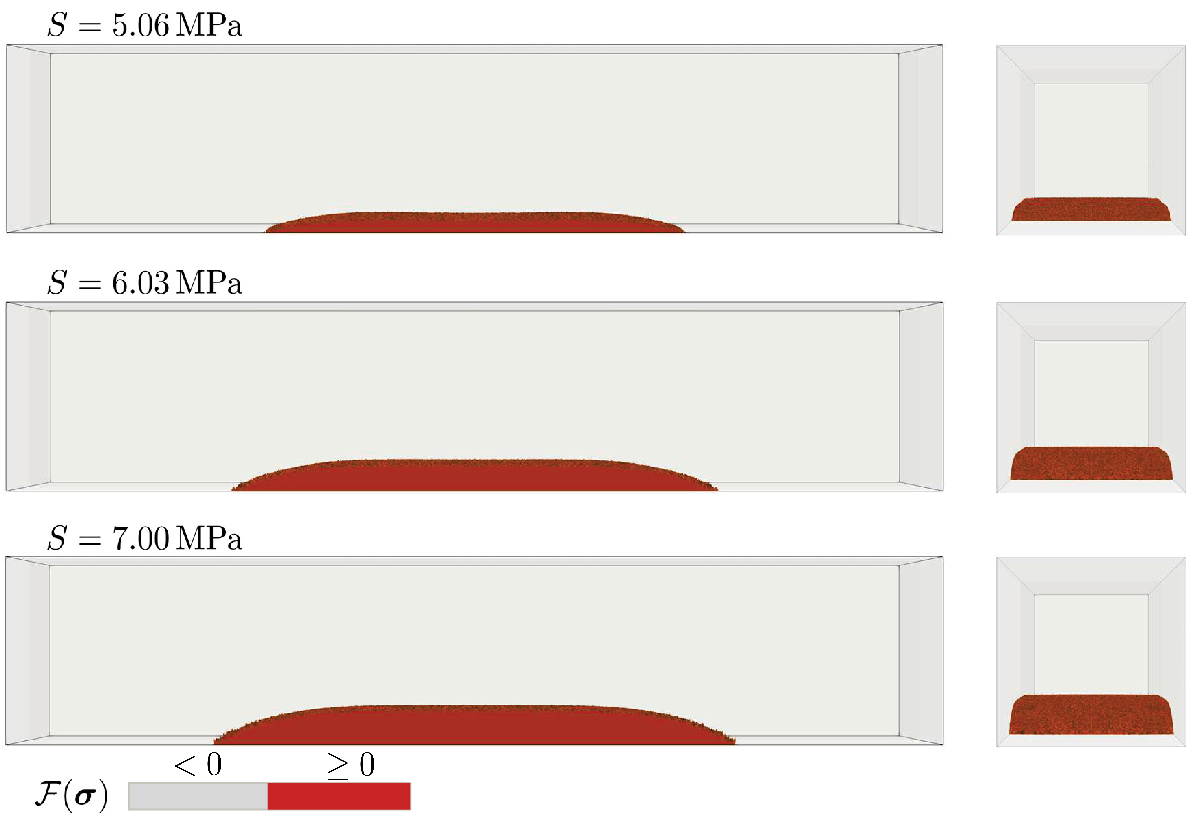}
\caption{{\small Contour plots --- shown from the $\bfe_1$-$\bfe_2$ (left) and the $\bfe_2$-$\bfe_3$ (right) perspectives --- of the regions (\ref{VF}) of the beam under four-point bending where the stress field exceeds ($\mathcal{F}(\boldsymbol{\sigma}\geq 0$) the strength surface of the material at three increasing values of the global stress $S$. The results correspond to the mortar with compressive-to-tensile strength ratio $\scs/\sts=15$.}}\label{Fig2}
\end{figure}
%
%
\begin{figure}[t!]
\centering
\includegraphics[width=0.925\linewidth]{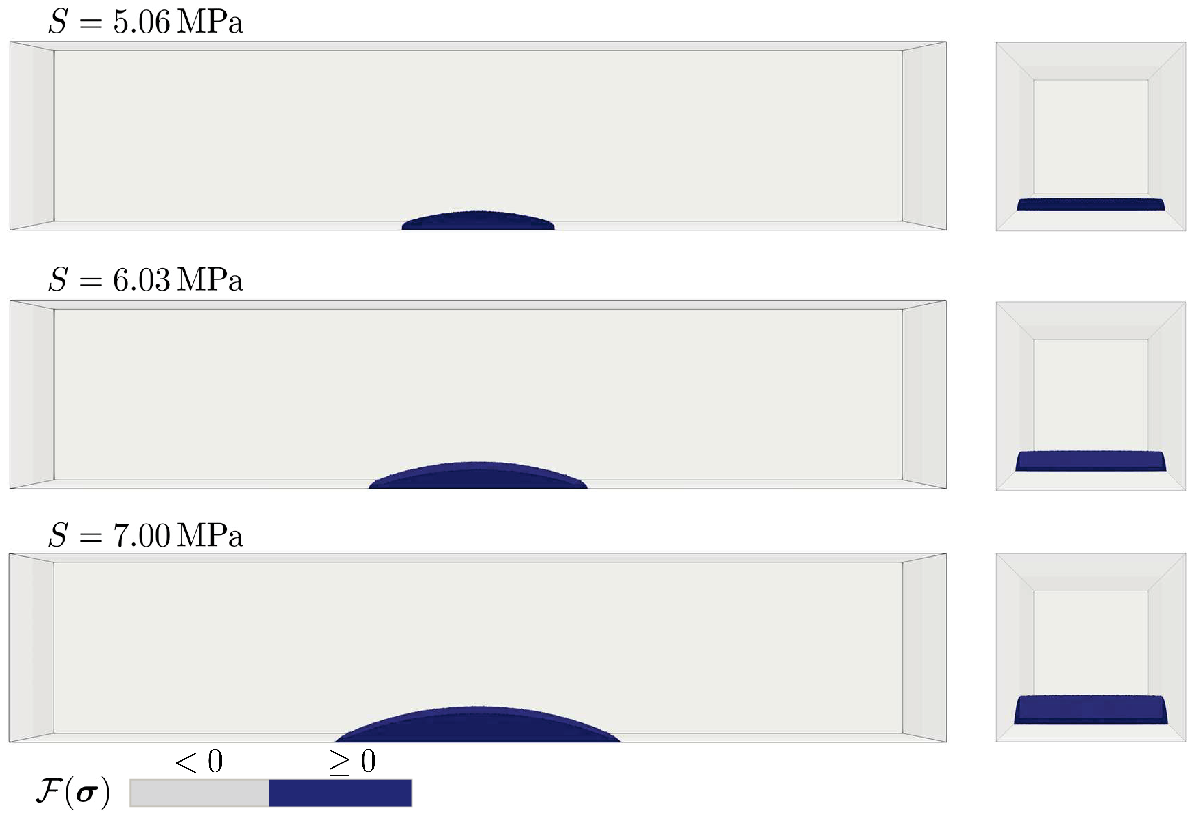}
\caption{{\small Contour plots --- shown from the $\bfe_1$-$\bfe_2$ (left) and the $\bfe_2$-$\bfe_3$ (right) perspectives --- of the regions (\ref{VF}) of the beam under three-point bending where the stress field exceeds ($\mathcal{F}(\boldsymbol{\sigma}\geq 0$) the strength surface of the material at three increasing values of the global stress $S$. The results correspond to the mortar with compressive-to-tensile strength ratio $\scs/\sts=15$.}}\label{Fig3}
\end{figure}
%

%
\begin{figure}[t!]
\centering
\includegraphics[width=0.68\linewidth]{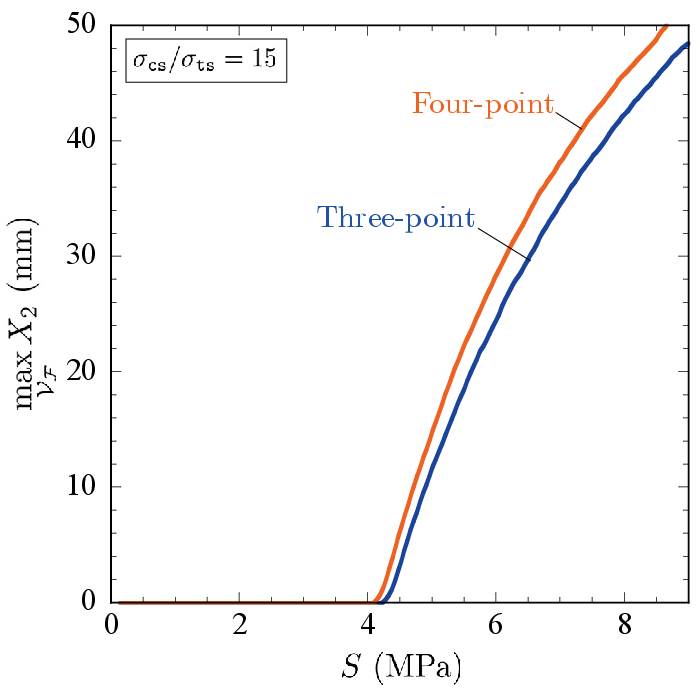}
\caption{{\small The maximum value of the $X_2$ coordinate in the strength violation set (\ref{VF}) shown in Figs.~\ref{Fig2} and \ref{Fig3}, as a function of the global stress $S$.}}\label{Fig4}
\end{figure}
%

Figures \ref{Fig2} and \ref{Fig3} present contour plots of the strength violation set (\ref{VF}) in the beam at three representative increasing values of the global stress $S$ for the four-point and three-point bending tests, respectively. The results are shown over the undeformed configuration from two different perspectives and pertain to a mortar with compressive-to-tensile strength ratio $\scs/\sts=15$. Figure \ref{Fig4} presents corresponding results for the maximum value of the $X_2$ coordinate in the strength violation set (\ref{VF}) --- that is, the maximum height at which the strength is exceeded --- as a function of $S$ for both bending tests. Several key observations can be made regarding these results. 

Starting with the \emph{when}, the strength surface is first violated at $S=4.15$ MPa in the beam under four-point bending and at $S=4.28$ MPa in the beam under three-point bending. These values are similar but \emph{not} the same as the uniaxial tensile strength $\sts=4$ MPa of the material, confirming the inexact nature of the Euler-Bernoulli approximation $\boldsymbol{\sigma}={\rm diag}(\sigma_1=-12M(X_2-H/2)/(DH^3),\sigma_2=0,\sigma_3=0)$ for the principal stresses. In particular, neither the maximum principal stress is given exactly by the formula $\sigma_1=-12M(X_2-H/2)/(DH^3)$ nor the two other stresses are exactly zero. More importantly, these values indicate that the strength of the material is exceeded earlier under four-point bending than under three-point bending. These results are consistent with experimental observations, in the sense that they hint at fracture occurring at lower loads (moments) under four-point bending than under three-point bending. In Section \ref{Sec: Results} below, we show that this is indeed the case.

\emph{Where} the strength surface is violated is also different for both tests. Under four-point bending, the violation occurs at the bottom of the beam, across its depth, over a wide region of the span. Under three-point bending, the violation also occurs at the bottom of the beam across its depth, but over a much more localized region (centered around $X_1=0$) within its span. 

As the load increases, the regions of strength violation grow in height and width in a fairly self-similar manner for both types of bending. As shown by  Fig.~\ref{Fig4}, this growth is slightly faster under four-point bending. This result too is consistent with the experimental observations that fracture is likely to occur at lower loads (moments) under four-point bending than under three-point bending.

\subsubsection{The effect of the compressive-to-tensile strength ratio $\sigma_{\texttt{\emph{cs}}}/\sigma_{\texttt{\emph{ts}}}$}

%
\begin{figure}[t!]
\centering
\includegraphics[width=0.68\linewidth]{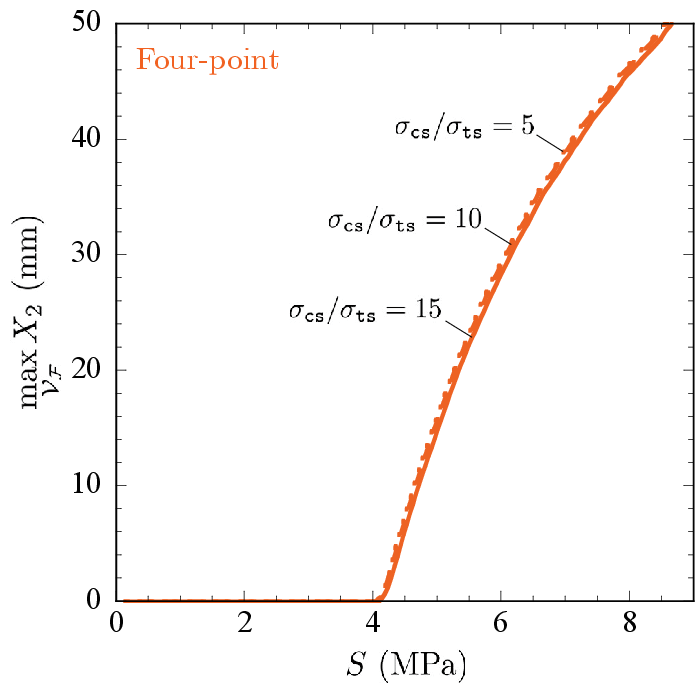}
\caption{{\small The maximum value of the $X_2$ coordinate in the strength violation set (\ref{VF}) in the beam under four-point bending, as a function of the global stress $S$. The results correspond to the mortars with compressive-to-tensile strength ratios $\scs/\sts=5,10,15$.}}\label{Fig5}
\end{figure}
%

Due to the linearity of the governing equations for the displacement field $\bfu(\bfX,t)$, the results presented above are qualitatively independent of the Young's modulus $E$ of the material. However, they may depend on its Poisson's ratio $\nu$, the ratio of compressive-to-tensile strength ratio $\scs/\sts$, and, more generally, on the form of its strength surface. 

\begin{figure}[t!]
\centering
\includegraphics[width=0.68\linewidth]{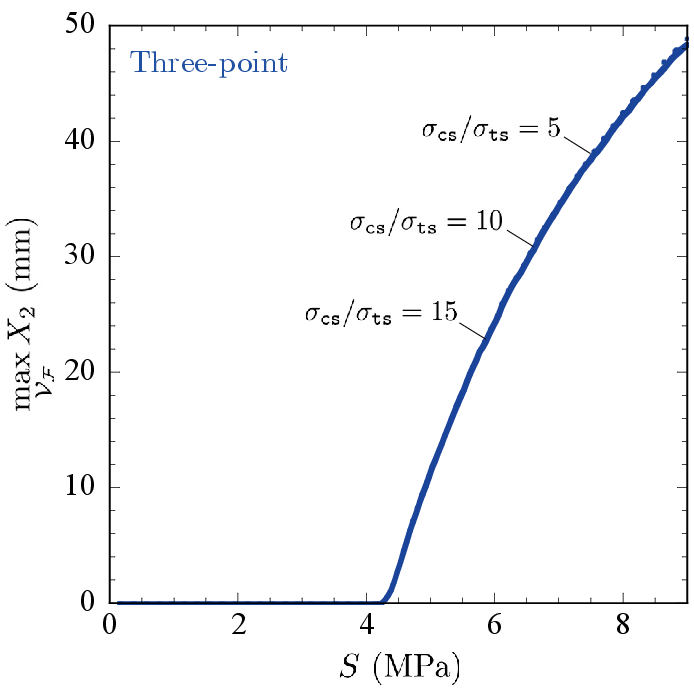}
\caption{{\small The maximum value of the $X_2$ coordinate in the strength violation set (\ref{VF}) in the beam under three-point bending, as a function of the global stress $S$. The results correspond to the mortars with compressive-to-tensile strength ratios $\scs/\sts=5,10,15$.}}\label{Fig6}
\end{figure}

As one may expect from basic intuition, simulations (not included here) for $\nu=0.10$ and $0.30$ (and the same beam geometry and remaining material constants) confirm that the results are only weakly dependent on the Poisson's ratio. Figures \ref{Fig5} and \ref{Fig6} illustrate that they are also largely insensitive to the compressive-to-tensile strength ratio $\scs/\sts$. In particular, these figures show results for the maximum value of the $X_2$ coordinate in the strength violation set (\ref{VF}), as a function of $S$, for $\scs/\sts=5,10,15$.  This insensitivity to $\scs/\sts$ is a manifestation of the fact that the principal stresses are such that $\sigma_1\gg \sigma_2,\sigma_3$ point-wise throughout the span away from the supports, which implies that the part of the strength surface that is violated is localized around $\boldsymbol{\sigma}={\rm diag}(\sigma_1\geq\sts,0,0)$. For this same reason, it is expected that the results are also insensitive to the form of the strength surface, not just the value of $\scs$. In other words, the specific type of strength surface $\mathcal{F}(\boldsymbol{\sigma})=0$ for the material of the interest --- whether is well described by a Drucker-Prager criterion, as assumed in this work, or any other criterion --- is expected to have little consequence on the nucleation of fracture in four-point and three-point bending tests.

%
%

\section{A complete fracture nucleation and propagation analysis of the bending tests}\label{Sec: Fracture}

In addition to the displacement $\bfu(\bfX,t)$, the boundary conditions applied in the bending tests eventually result in the nucleation and subsequent propagation of cracks in the beam. We represent these cracks using a regularized approach, employing an order parameter or phase field:

\begin{equation*}
v=v(\bfX, t).
\end{equation*}
This field is constrained to the interval $[0,1]$, where $v=1$ represents intact material and $v=0$ indicates fully fractured regions. The transition between these states occurs smoothly across a localized zone of length scale $\varepsilon>0$.

\subsection{The governing equations of elastic deformation and fracture}

According to the specialization of the phase-field fracture theory \citep*{KFLP18} to the four-point and three-point bending tests of interest in this work, the displacement field $\bfu_k(\bfX)=\bfu(\bfX,t_k)$ and phase field $v_k(\bfX)=v(\bfX,t_k)$ at any material point $\bfX \in \overline{\Omega}=\Omega\cup\partial\Omega$ and at any given discrete time $t_k\in\{0=t_0,t_1,...,t_m,t_{m+1},...,t_M=T\}$ are determined by the system of coupled partial differential equations (PDEs)
\begin{equation}
\left\{\begin{array}{ll}
\hspace{-0.15cm} {\rm Div}\left[v_{k}^2 \boldsymbol{\sigma}(\bfX,t_k)\right]=\textbf{0},& \bfX\in\Omega\vspace{0.2cm}\\
\hspace{-0.15cm} {\rm Div}\left[\varepsilon\, \delta^\varepsilon G_c \nabla v_k\right]=\dfrac{8}{3}v_{k} W(\bfE(\bfu_k))+\vspace{0.1cm}\\
\hspace{-0.15cm}\dfrac{4}{3}c_\texttt{e}(\bfX,t_{k})-\dfrac{\delta^\varepsilon G_c}{2\varepsilon}+\dfrac{8}{3\,\zeta} \, p(v_{k-1},v_k),& \bfX\in \Omega
\end{array}\right. \label{PDEs-PF}
\end{equation}
subject to the boundary conditions
\begin{equation}
\left\{\begin{array}{ll}
\hspace{-0.2cm} \left\{\begin{array}{l}
\hspace{-0.15cm}v_{k}^2\sigma_{1j}(\bfX,t_k)N_j=0\vspace{0.2cm}\\
\hspace{-0.15cm}u_2(\bfX,t_k)=0\vspace{0.2cm}\\
\hspace{-0.15cm}v_{k}^2\sigma_{3j}(\bfX,t_k)N_j=0
\end{array}\right. \hspace{-0.15cm}, &\hspace{-0.05cm} \bfX\in\partial\Omega_{\mathcal{B}}\vspace{0.2cm}\\
\hspace{-0.2cm} \left\{\begin{array}{l}
\hspace{-0.15cm}v_{k}^2\sigma_{1j}(\bfX,t_k)N_j=0\vspace{0.2cm}\\
\hspace{-0.15cm}u_2(\bfX,t_k)=-u\vspace{0.2cm}\\
\hspace{-0.15cm}v_{k}^2\sigma_{3j}(\bfX,t_k)N_j=0
\end{array}\right. \hspace{-0.15cm}, &\hspace{-0.05cm} \bfX\in\partial\Omega_{\mathcal{T}}\vspace{0.2cm}\\
\hspace{-0.15cm}v_{k}^2 \boldsymbol{\sigma}(\bfX,t_k)\bfN=\textbf{0},& \bfX\in\partial\Omega\setminus\partial\Omega_{\mathcal{B}}\cup\partial\Omega_{\mathcal{T}}\vspace{0.2cm}\\
\hspace{-0.15cm} \nabla v_k\cdot\bfN=0,& \bfX\in \partial\Omega
\end{array}\right. \label{BCs-PF}
\end{equation}
(no summation in $k$), where $\partial\Omega_{\mathcal{T}}=\partial\Omega^{4p}_{\mathcal{T}}$ and $\partial\Omega_{\mathcal{T}}=\partial\Omega^{3p}_{\mathcal{T}}$ for the four-point bending and the three-point bending tests, respectively. In these equations, we recall that the stress $\boldsymbol{\sigma}(\bfX,t)$ and stored-energy function $W(\bfE(\bfu))$ are given in terms of the displacement field $\bfu(\bfX,t)$ by relations (\ref{S-E-nu}) and (\ref{W-E-nu}). The penalty function $p(v_{k-1},v_k)=|v_{k-1}-v_k|-(v_{k-1}-v_k)-|v_k|+v_k$ and penalty parameter\footnote{Typically, it suffices to set $\zeta^{-1}=10^4 \delta^\varepsilon G_c/(2\varepsilon)$.} $\zeta$ enforce that the phase field remains in the physically admissible range $0\leq v\leq 1$ and that fracture is irreversible. Making use of the constitutive prescription\footnote{The constitutive prescription for $c_{\texttt{e}}$ depends on the particular form of the strength surface $\mathcal{F}(\bfS)=0$ of the material. For the case of the Drucker-Prager strength surface (\ref{DP-1}) of interest here, it is given by (\ref{ce}). For other strength surfaces, corresponding prescriptions for $c_{\texttt{e}}$ can be constructed by following the blueprint outlined by \cite{KLP20} and \cite{KBFLP20}; see, e.g., \citep{Chockalingam25}.} in \citep{KKLP24}, 
\begin{align}
c_{\texttt{e}}(\bfX,t)&=v^2\beta_2\sqrt{J_2}+v^2\beta_1 I_1-v\left(1-\dfrac{\sqrt{I^2_1}}{I_1}\right)W(\bfE(\bfu))\label{ce}
\end{align}
with 
\begin{equation}\label{deltaeps}
\left\{\hspace{-0.15cm}\begin{array}{l}
\beta_1=\dfrac{1}{\shs}\delta^\varepsilon\dfrac{G_c}{8\varepsilon}-\dfrac{2 W_{\texttt{hs}}}{3\shs}\vspace{0.2cm}\\
\beta_2=\dfrac{\sqrt{3}(3\shs-\sts)}{\shs\sts}\delta^\varepsilon\dfrac{G_c}{8\varepsilon}+
\dfrac{2W_{\texttt{hs}}}{\sqrt{3}\shs}-\dfrac{2\sqrt{3}W_{\texttt{ts}}}{\sts}\vspace{0.2cm}\\
\delta^\varepsilon=\left(\dfrac{\sts+(1+2\sqrt{3})\,\shs}{(8+3\sqrt{3})\,\shs}\right)\dfrac{3 G_c}{16 W_{\texttt{ts}}\varepsilon}+\dfrac{2}{5}
\end{array}\right.,
\end{equation}
$W_{\texttt{ts}}=\sts^2/(2E)$, $W_{\texttt{hs}}=3(1-2\nu)\shs^2/(2 E)$, and $\shs=2 \scs\sts/(3(\scs-\sts))$.

The governing equations (\ref{PDEs-PF})-(\ref{BCs-PF}) can be solved numerically via FEs. Open-source implementations are available, for instance, in FEniCS\footnote{\url{https://pamies.cee.illinois.edu/repositories/}} and MOOSE.\footnote{\url{https://github.com/hugary1995/raccoon.}} In this work, we make use of the FEniCS implementation. All simulations presented below pertain to values $\varepsilon\in[1.5, 4]$ mm for the regularization length, which are small enough\footnote{Typically, it suffices to set $\varepsilon< 3 G_c/(16 W_{\texttt{ts}})$; see, e.g., Remark 6 in \citep{KKLP24}.} for the beam under study here, and make use of unstructured linear tetrahedral meshes of minimum sizes $\texttt{h}=\varepsilon/5\in[0.3,0.8]$ mm.

\begin{remark}
\emph{When using the FE method to solve the governing equations (\ref{PDEs-PF})-(\ref{BCs-PF}), meshes of small enough element size $\texttt{h}$ ought to be used so as to appropriately resolve the spatial variations of the phase field $v_k$ over lengths of order $\varepsilon$. Nevertheless, an error is incurred that scales with $\texttt{h}$. It is possible to include a correction in the formula (\ref{deltaeps})$_3$ for $\delta^\varepsilon$ so that the FE solutions of equations (\ref{PDEs-PF})-(\ref{BCs-PF}) are consistent with the actual value $G_c$ of the critical energy release rate of the material. For first-order FEs of size $\texttt{h}$, the formula for $\delta^\varepsilon$ with the correction reads \citep{KKLP24}
\begin{align*}
\delta^\varepsilon=&\left(1+\dfrac{3}{8}\dfrac{\texttt{h}}{\varepsilon}\right)^{-2}\left(\dfrac{\sts+(1+2\sqrt{3})\,\shs}{(8+3\sqrt{3})\,\shs}\right)\dfrac{3 G_c}{16 W_{\texttt{ts}}\varepsilon}+\\
&\left(1+\dfrac{3}{8}\dfrac{\texttt{h}}{\varepsilon}\right)^{-1}\dfrac{2}{5}.
\label{delta-eps-final-h}
\end{align*}
}
\end{remark}

\section{Results and discussion}\label{Sec: Results}

Having spelled out the governing equations (\ref{PDEs-PF})-(\ref{BCs-PF}), we are now in a position to investigate in a precise and quantitative manner when and where cracks nucleate and propagate in the beam under four-point and three-point bending.

We begin in Subsection \ref{SubS-4p-3p} by confronting the response of the beam under four-point bending with that under three-point bending for one of the mortars with the material constants listed in Table \ref{Table1}. Subsequent subsections examine the dependence of these results on four key material properties: Subsections \ref{SubS-scssts} and \ref{SubS-Gc} explore the influence of the compressive-to-tensile strength ratio $\scs/\sts$ and critical energy release rate $G_c$, respectively, while Subsections \ref{SubS-lc} and \ref{SubS-Stochasticity} investigate the relevance of the material length scale $\ell_{\texttt{ts}}:=EG_c/\sts^2$ and how incorporating strength stochasticity modifies the response under both loading conditions.

\subsection{Four-point vs. three-point bending}\label{SubS-4p-3p}

%
\begin{figure}[b!]
\centering
\includegraphics[width=0.68\linewidth]{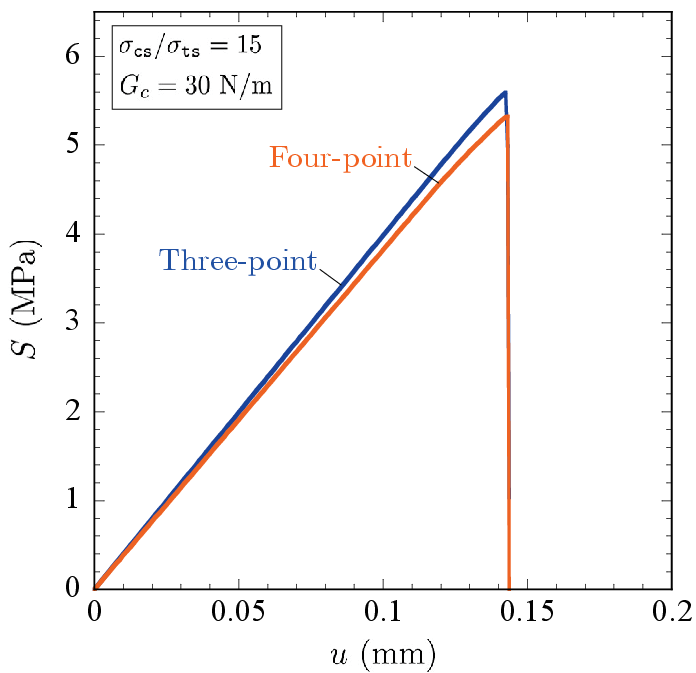}
\caption{{\small Comparison of the responses of the beam under four-point bending and three-point bending. The results show the global stress $S$ as a function of the applied displacement $u$ for the mortar with compressive-to-tensile strength ratio $\scs/\sts=15$ and critical energy release rate $G_c=30$ N/m.}}\label{Fig7}
\end{figure}
%
%
\begin{figure}[t!]
\centering
\includegraphics[width=0.85\linewidth]{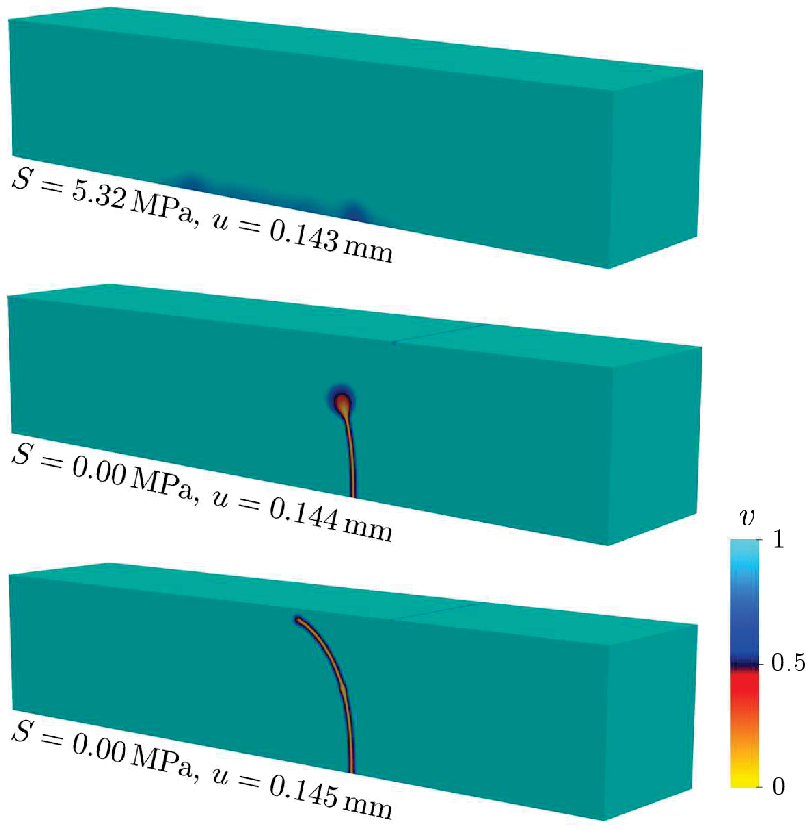}
\caption{{\small Contour plots of the phase field $v$ in the beam under four-point bending at three pairs of values of the global stress $S$ and applied displacement $u$.}}\label{Fig8}
\end{figure}
%
%
\begin{figure}[t!]
\centering
\includegraphics[width=0.85\linewidth]{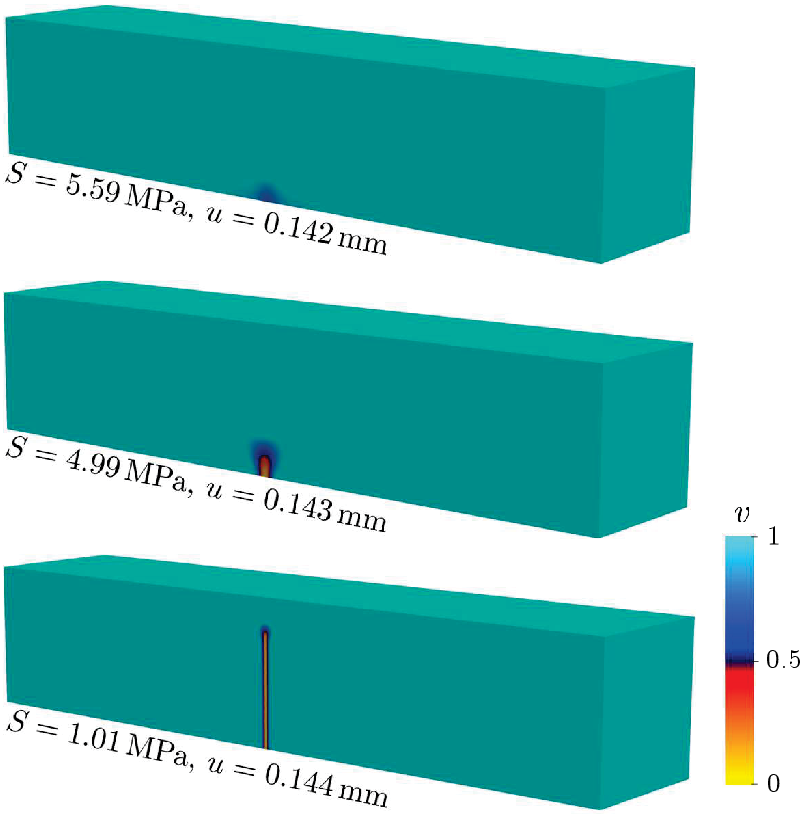}
\caption{{\small Contour plots of the phase field $v$ in the beam under three-point bending at three pairs of values of the global stress $S$ and applied displacement $u$.}}\label{Fig9}
\end{figure}
%

Figure \ref{Fig7} compares the global responses of the beam under four-point bending and three-point bending, from its initial deformation all the way until it is severed into two pieces at sufficiently large loads. In particular, the figure shows plots of the global stress $S$ as a function of the applied displacement $u$ for the mortar with compressive-to-tensile strength ratio $\scs/\sts=15$ and critical energy release rate $G_c=30$ N/m; see Table \ref{Table1}. Figures \ref{Fig8} and \ref{Fig9} present associated contour plots (over the undeformed configuration) of the phase field $v$ at three different values of the pair $(S,u)$ showing its evolution, that is, the process of crack nucleation and subsequent propagation. There are three main observations to be made from these results.

First, in agreement with experimental observations, Fig.~\ref{Fig7} shows that the global stress $S$ increases monotonically with increasing displacement $u$ until reaching a local maximum --- the flexural strength $S_{max}$. Beyond this point, the value of the global stress suddenly drops to a small value for both types of loading. As shown by Figs.~\ref{Fig8} and \ref{Fig9}, the sudden drop in $S$ is the manifestation of the brutal nucleation of a finite-size crack at the bottom of the beam that subsequently propagates in a very rapid manner eventually severing the beam.

Second, the crack topologies differ significantly between the two types of loading. In the four-point bending test, the crack nucleates off-center between the inner supports, at around $X_1=136$ mm in this particular case, and curves toward the midspan rather than propagating straight ahead. In contrast, under three-point bending, the crack nucleates and propagates straight along the center of the beam. Both of these behaviors are consistent with experimental observations. Importantly, while the four-point bending setup is geometrically and mechanically symmetric, the resulting crack is not. This asymmetry is triggered by small numerical noise within the FE solution, which is sufficient to break the symmetry of the system. This phenomenon is explored further in Subsection \ref{SubS-Stochasticity} below, where we relax the idealization that the material strength is spatially uniform throughout the beam and explicitly account for its stochasticity.

Third, the maximum global stress at which fracture nucleation occurs is smaller for the beam under four-point bending, $S^{4p}_{max}=5.32$ MPa, than under three-point bending, $S^{3p}_{max}=5.59$ MPa, in particular, $(S^{3p}_{max}-S^{4p}_{max})/$ $S^{3p}_{max}=5\%$. This is consistent with the preliminary strength analysis presented in Section \ref{Sec: Strength} above and, more importantly, consistent with experimental observations. In some experiments, however, the differences can be larger, between $10\%$ and $20\%$; see, e.g., \citep{Wright1952,Lindner1955} for classical experiments on concrete. As we elaborate in Subsection \ref{SubS-Stochasticity} and Section \ref{Sec: Final Comments}, these larger differences can be attributed to strength stochasticity, different beam dimensions ($L_s/H$, $D/H$, $H$), and different loading-span-to-height ratios ($L_l/H$).

\subsection{The effect of the compressive-to-tensile strength ratio $\sigma_{\texttt{\emph{cs}}}/\sigma_{\texttt{\emph{ts}}}$}\label{SubS-scssts}

\begin{figure}[b!]
\centering
\includegraphics[width=0.68\linewidth]{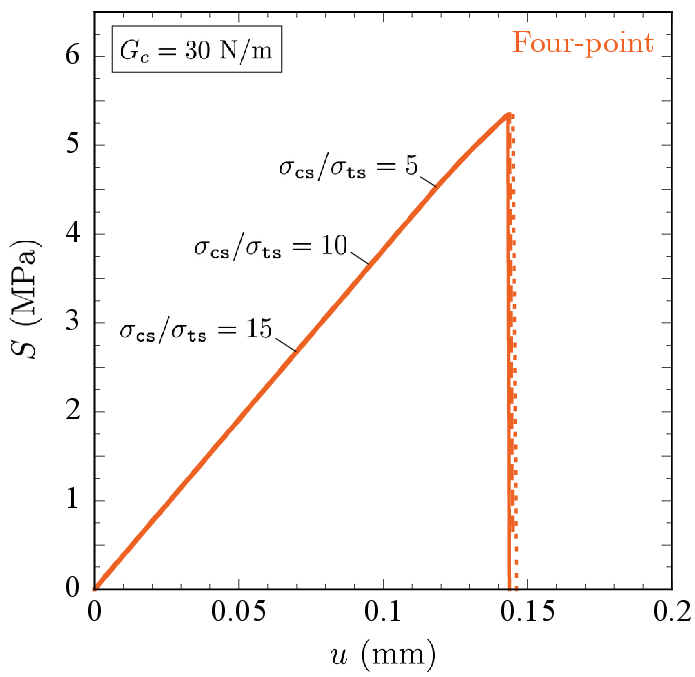}
\caption{{\small The global response ($S$ vs. $u$) of the beam under four-point bending for three different mortars with critical energy release rate $G_c=30$ N/m and compressive-to-tensile strength ratios $\scs/\sts=5, 10, 15$.}}\label{Fig10}
\end{figure}

\begin{figure}[t!]
\centering
\includegraphics[width=0.68\linewidth]{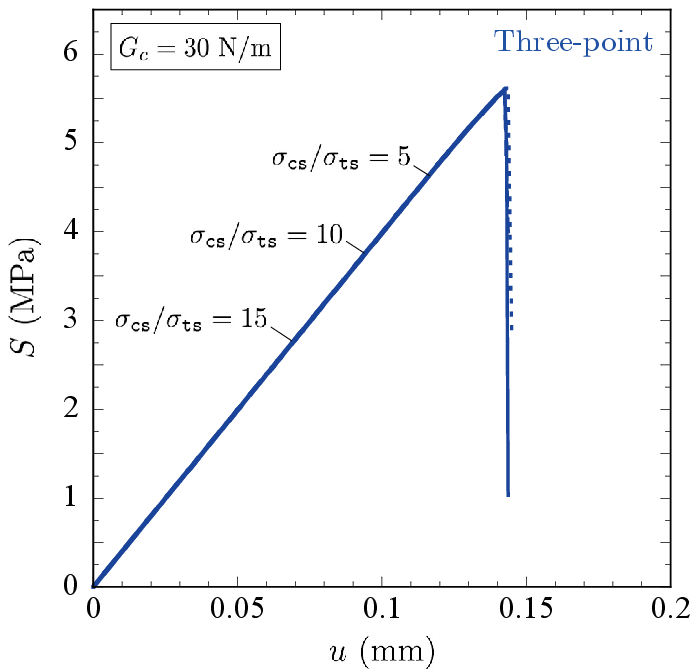}
\caption{{\small The global response ($S$ vs. $u$) of the beam under three-point bending for three different mortars with critical energy release rate $G_c=30$ N/m and compressive-to-tensile strength ratios $\scs/\sts=5, 10, 15$.}}\label{Fig11}
\end{figure}

We next examine the sensitivity of the response of the beam to the compressive-to-tensile strength ratio $\scs/\sts$. To that end, Figs.~\ref{Fig10} and \ref{Fig11} present results for the global stress $S$ versus the applied displacement $u$ for four-point and three-point bending tests, respectively, for mortars with critical energy release rate $G_c=30$ N/m and three different compressive-to-tensile strength ratios: $\scs/\sts=5, 10, 15$.
%
%
%
%

Consistent with the preliminary strength analysis presented in Section \ref{Sec: Strength} above, the main observation from these results is that the uniaxial compressive strength $\scs$ plays essentially no role on when and where cracks nucleate under either type of loading. 

The practical implication of this insensitivity is significant: it indicates that the only part of the strength surface $\mathcal{F}(\boldsymbol{\sigma})=0$ that is effectively engaged in the nucleation of fracture in the beam under both types of bending is the uniaxial tensile strength $\sts$. Consequently, we should be able to derive a direct relation between the maximum global stress $S_{max}$ observed experimentally and the uniaxial tensile strength $\sts$ of the material being tested irrespective of how the strength surface $\mathcal{F}(\boldsymbol{\sigma})=0$ away from uniaxial tension $\boldsymbol{\sigma}={\rm diag}(\sigma>0,0,0)$  looks like. We do just that in Section \ref{Sec: Final Comments} below.

\subsection{The effect of the critical energy release rate $G_c$}\label{SubS-Gc}

%
\begin{figure}[b!]
\centering
\includegraphics[width=0.68\linewidth]{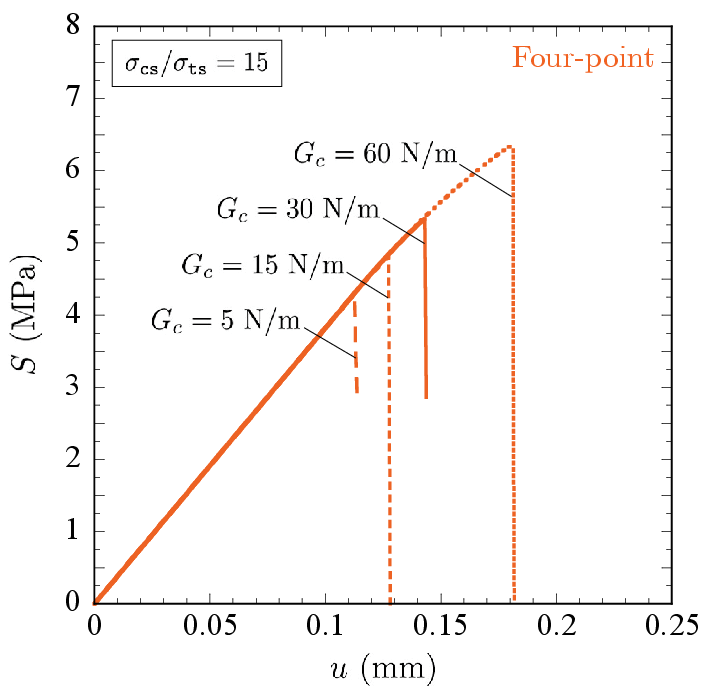}
\caption{{\small The global response ($S$ vs. $u$) of the beam under four-point bending for three different mortars with compressive-to-tensile strength ratio $\scs/\sts=15$ and critical energy release rates $G_c=5, 15, 30, 60$ N/m.}}\label{Fig12}
\end{figure}
%

%
\begin{figure}[t!]
\centering
\includegraphics[width=0.68\linewidth]{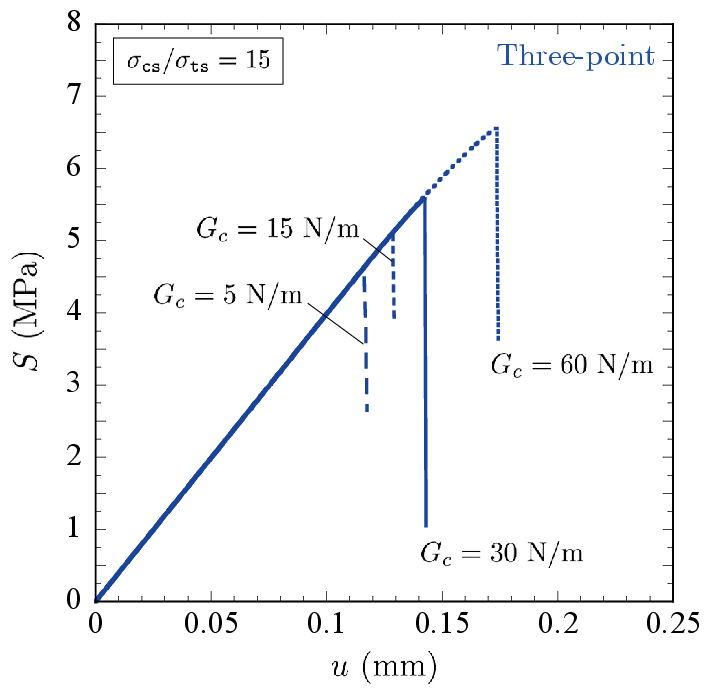}
\caption{{\small The global response ($S$ vs. $u$) of the beam under three-point bending for three different mortars with compressive-to-tensile strength ratio $\scs/\sts=15$ and critical energy release rates $G_c=5, 15, 30, 60$ N/m.}}\label{Fig13}
\end{figure}
%

%
\begin{figure}[t!]
\centering
\includegraphics[width=0.68\linewidth]{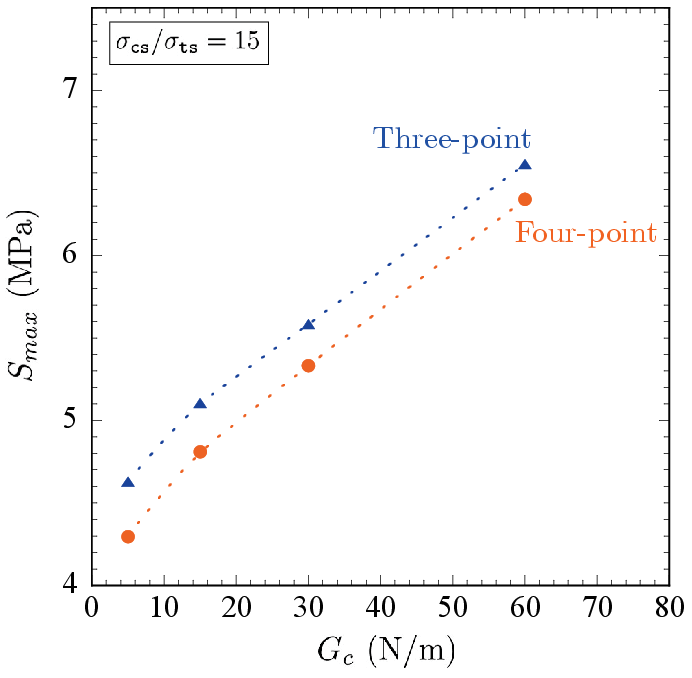}
\caption{{\small The maximum global stress, $S_{max}^{4p}$ and $S_{max}^{3p}$, at which fracture nucleates in the beam under four-point and three-point bending, as a function of the critical energy release rate $G_c$. The results pertain to mortars with compressive-to-tensile strength ratio $\scs/\sts=15$.}}\label{Fig14}
\end{figure}
%

We now explore the effect of the critical energy release rate $G_c$. Analogous to the preceding subsection, Figs.~\ref{Fig12} and \ref{Fig13} present results for the global stress $S$ versus the applied displacement $u$ for four-point and three-point bending tests, respectively, for mortars with compressive-to-tensile strength ratio $\scs/\sts=15$ and four different critical energy release rates: $G_c=5, 15, 30, 60$ N/m.

For both types of bending, the main observation from these results is that larger values of $G_c$ lead to significantly larger values of the maximum global stress, $S_{max}^{4p}$ and $S_{max}^{3p}$, at which fracture nucleation occurs. This strong dependence is better illustrated by Fig.~\ref{Fig14}, where $S_{max}^{4p}$ and $S_{max}^{3p}$ are plotted as functions of $G_c$. Notably, the maximum global stress under three-point bending consistently exceeds that under four-point bending, with both remaining higher than the material's uniaxial tensile strength across the entire range of $G_c$ values: $S_{max}^{3p}>S_{max}^{4p}>\sts$.

A practical implication of this dependence is that any formula that aims at relating the uniaxial tensile strength $\sts$ of the material to the maximum global stress measured from either bending test must necessarily also involve the critical energy release rate $G_c$. In other words, the values $S_{max}^{4p}$ and $S_{max}^{3p}$ that are observed experimentally depend \emph{not} only on $\sts$, but also on $G_c$. We elaborate on this key dependence next.

\subsection{The dependence of $S_{max}$ on $E$, $\sigma_{\emph{\texttt{ts}}}$, and $G_c$}\label{SubS-lc}

In terms of material constants, the results presented up to this point have established that the maximum global stress $S_{max}$ at which fracture nucleates in the beam under either type of bending depends primarily on the Young's modulus $E$, the uniaxial tensile strength $\sts$, and the critical energy release rate $G_c$. Notably, $S_{max}$ is practically independent of the Poisson's ratio $\nu$ and the topology of the strength surface $\mathcal{F}(\boldsymbol{\sigma})=0$ outside uniaxial tension.

The results have also established that fracture nucleation only occurs when the strength surface $\mathcal{F}(\boldsymbol{\sigma})=0$ of the material is exceeded over a region of sufficient height --- a material-specific length --- such that forming a crack within that region is energetically favorable. 

Collectively, these findings suggest that the dependence of $S_{max}$ on $E$, $\sts$, and $G_c$ may be primarily, if not exclusively, through two combinations of these material constants: the uniaxial tensile strength itself,
\begin{equation*}
\sts,
\end{equation*}
and the classical material length scale
\begin{equation*}
\ell_{\texttt{ts}}=\dfrac{E G_c}{\sts^2}.
\end{equation*}
More precisely, the results suggests that the maximum global stress at which fracture nucleates in the beam can be expressed as a function of these two material properties, in addition, of course, of its geometric dimensions ($L_s,D,H$) and loading configuration ($L_l$):
\begin{equation}\label{Smax-0}
S_{max}=S_{max}\left(\dfrac{L_s}{H},\dfrac{D}{H},H,\dfrac{L_l}{H},\sts,\dfrac{E G_c}{\sts^2}\right).
\end{equation}
\begin{remark}
\emph{Recall that the calculations presented thus far and in the remaining of this section correspond to a fixed beam geometry with $L_s/H=4$, $D/H=1$, and $H=0.2$ m, while the loading configurations correspond to $L_l/H=2$ for four-point bending and $L_l/H=0$ for three-point bending. We shall consider additional beam sizes in Section \ref{Sec: Final Comments}.}
\end{remark}

To verify the functional dependencies proposed in (\ref{Smax-0}) on $E$, $\sts$, and $G_c$, we conducted simulations across a range of values for $\sts$ and $G_c$  while maintaining a constant value for the material length scale $E G_c/\sts^2$. Note that varying $E$ independently is unnecessary when already varying $\sts$ and so the Young's modulus was kept at $E=27$ GPa in all the simulations. Representative results (calculated with $\nu=0.20$ and $\scs=60$ MPa) for four-point and three-point bending are presented in Tables \ref{Table2} and \ref{Table3}, respectively. These data\footnote{In passing, for the three sets of material lengths scales in Tables \ref{Table2} and \ref{Table3}, it is worth remarking that classical variational phase-field models, such as the AT$_1$ and the star-convex models \citep{Vicentini24}, would require the use of unphysically large regularization lengths: $\varepsilon=3 E G_c/(8\sts^2)=9.5, 19,$ and $38$ mm.}  not only verify the dependencies on $E$, $\sts$, and $G_c$, but further reveal that the maximum global stress follows the more specific form:
\begin{equation*}
S_{max}=g\left(\dfrac{L_s}{H},\dfrac{D}{H},\dfrac{L_l}{H},\dfrac{E G_c}{H\sts^2}\right)\sts,
\end{equation*}
where the dimensionless function $g$ is bounded from below by 1 and increases monotonically with respect to its final argument, the normalized material length scale $E G_c/(H\sts^2)$. As elaborated in \ref{Sec: Final Comments} below, it is possible to fit the predictions for $S_{max}$ with simple explicit formulas for $g$. 

\begin{table}[H]\centering\small
\caption{Values predicted by the phase-field theory for the maximum global stress $S^{4p}_{max}$ at which fracture nucleates in the beam under four-point bending, calculated for various $\sts$ and $G_c$ values at several constant material length scales $E G_c/\sts^2$.}
\begin{tabular}{ccc?cc}
\toprule
$\sts$                    & $G_c$                     &  $E G_c/\sts^2$  & $S^{4p}_{max}$       &  $S^{4p}_{max}/\sts$      \\
\footnotesize{(MPa)}      & \footnotesize{(N/m)}      &  \footnotesize{(mm)}                   & \footnotesize{(MPa)} &   \\
\midrule
\midrule
4            & 15        & 25.3                      & 4.812  &   1.203         \\
\midrule
5.657        & 30        & 25.3                       & 6.794     &  1.201         \\
\midrule
\midrule
4            & 30        & 50.6                      & 5.324    &   1.331         \\
\midrule
1.633          & 5        & 50.6                     & 2.187     &   1.339        \\
\midrule
\midrule
4           & 60        & 101.3                        & 6.288      &  1.572      \\
\midrule
2.828            & 30        & 101.3                        & 4.491   & 1.588            \\
\bottomrule
\end{tabular} \label{Table2}
\end{table}
\begin{table}[H]\centering\small
\caption{Values predicted by the phase-field theory for the maximum global stress $S^{4p}_{max}$ at which fracture nucleates in the beam under three-point bending, calculated for various $\sts$ and $G_c$ values at several constant material length scales $E G_c/\sts^2$.}
\begin{tabular}{ccc?cc}
\toprule
$\sts$                    & $G_c$                     &  $E G_c/\sts^2$  & $S^{3p}_{max}$       &  $S^{3p}_{max}/\sts$      \\
\footnotesize{(MPa)}      & \footnotesize{(N/m)}      &  \footnotesize{(mm)}                   & \footnotesize{(MPa)} &   \\
\midrule
\midrule
4            & 15        & 25.3                      & 5.104   &    1.276        \\
\midrule
5.657        & 30        & 25.3                       & 7.230   &    1.278       \\
\midrule
\midrule
4            & 30        & 50.6                      & 5.592    &    1.398        \\
\midrule
1.633          & 5        & 50.6                     & 2.280     &   1.396        \\
\midrule
\midrule
4           & 60        & 101.3                        & 6.552      &  1.638       \\
\midrule
2.828            & 30        & 101.3                        & 4.638   &   1.640         \\
\bottomrule
\end{tabular} \label{Table3}
\end{table}

\subsection{The effect of strength stochasticity}\label{SubS-Stochasticity}

All the results that have been presented thus far  pertain to deterministic values for the beam dimensions, loading conditions, and material properties. In practice, as for any test, there is uncertainty and stochasticity associated with all such values; see, e.g., \citep{Wright1952,Hori59,Strange79,Le23,Weinberg25}. In this section, we study the effect of the spatial stochasticity of the strength --- which is arguably the primary source of variability in bending tests --- on the response of the beam. Before proceeding with the analysis of strength stochasticity \emph{per se}, however, we first need discuss the possible presence of a boundary layer within the beam.

\subsubsection{The presence of a boundary layer and its dominant role in fracture nucleation}

\begin{figure}[H]
\centering
\includegraphics[width=0.75\linewidth]{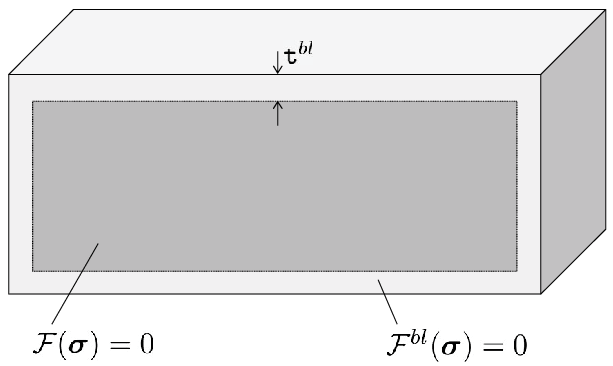}
\caption{{\small Schematic of a beam illustrating its boundary layer, of thickness $\texttt{t}^{bl}$, wherein the strength surface $\mathcal{F}^{bl}(\boldsymbol{\sigma})=0$ is possibly different from that ($\mathcal{F}(\boldsymbol{\sigma})=0$) in the bulk because the underlying defects are different due to the fabrication process. Throughout this section, we consider a boundary layer of thickness $\texttt{t}^{bl}=H/8=25$ mm.}}\label{Fig15}
\end{figure}

As already noted in Section \ref{Sec: Strength}, the strength surface $\mathcal{F}(\boldsymbol{\sigma})=0$ of a material is the macroscopic manifestation of its microscopic defects. In beams, fabrication processes often induce defects at and around the boundary that differ from those within the bulk. For instance, reaction kinetics at the mold interface during curing may diverge from the interior behavior. In cementitious materials like mortar and concrete, localized water loss at this interface can produce a cement paste that is weaker than the bulk, an effect often compounded by inefficient aggregate packing near the boundary. Similarly, mechanical cutting processes can introduce localized surface damage. Here, we account for the presence of such defects by assuming that there is a boundary layer, of thickness $\texttt{t}^{bl}=H/8=25$ mm, around the beam that features a Drucker-Prager strength surface
\begin{equation*}
\mathcal{F}^{bl}(\boldsymbol{\sigma})=\sqrt{J_2}+\dfrac{\sigma_{\texttt{cs}}-\sigma^{bl}_{\texttt{ts}}}
{\sqrt{3}\left(\sigma_{\texttt{cs}}+\sigma^{bl}_{\texttt{ts}}\right)} I_1-\dfrac{2\sigma_{\texttt{cs}}\sigma^{bl}_{\texttt{ts}}}
{\sqrt{3}\left(\sigma_{\texttt{cs}}+\sigma^{bl}_{\texttt{ts}}\right)}=0
\end{equation*}
with a possibly different uniaxial tensile strength\footnote{Recall that we have already established in Subsection \ref{SubS-scssts} that the only part of the strength surface $\mathcal{F}(\boldsymbol{\sigma})=0$ that is effectively engaged in the nucleation of fracture in the beam under both types of bending is the uniaxial tensile strength.}  $\sigma^{bl}_{\texttt{ts}}$ than that ($\sigma_{\texttt{ts}}$) in the strength surface (\ref{DP-1}) for the bulk; see Fig.~\ref{Fig15}. 
%
%

%
\begin{figure}[t!]
\centering
\includegraphics[width=0.68\linewidth]{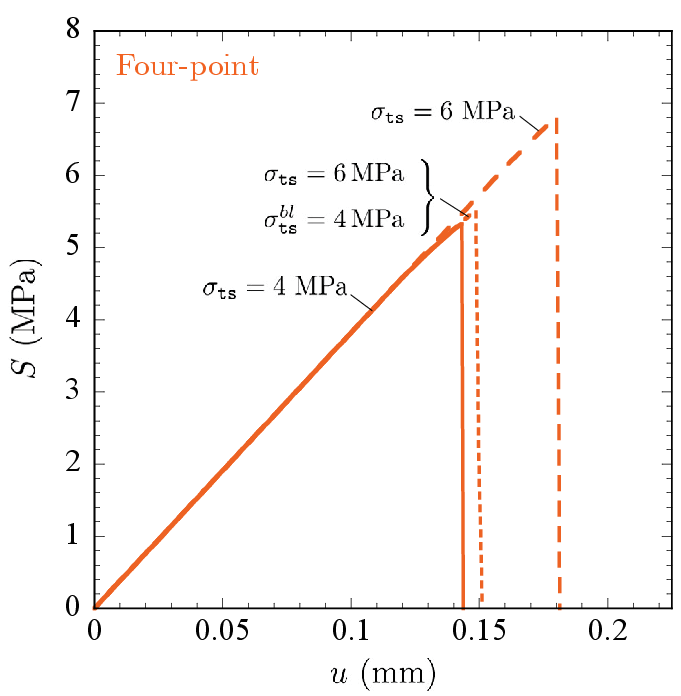}
\caption{{\small The global response ($S$ vs. $u$) of the beam under four-point bending for three mortars: two with different uniform uniaxial tensile strengths, $\sts=4$ MPa and $\sts=6$ MPa, and one with bulk uniaxial tensile strength $\sts=6$ MPa and a $25$-mm-thick boundary layer with uniaxial tensile strength $\sts^{bl}=4$ MPa. For all three mortars, the remaining constants are the same as in Fig.~\ref{Fig7}: $E=27$ GPa, $\nu=0.20$, $\scs=60$ MPa, and $G_c=30$ N/m.}}\label{Fig16}
\end{figure}
%
%
\begin{figure}[t!]
\centering
\includegraphics[width=0.68\linewidth]{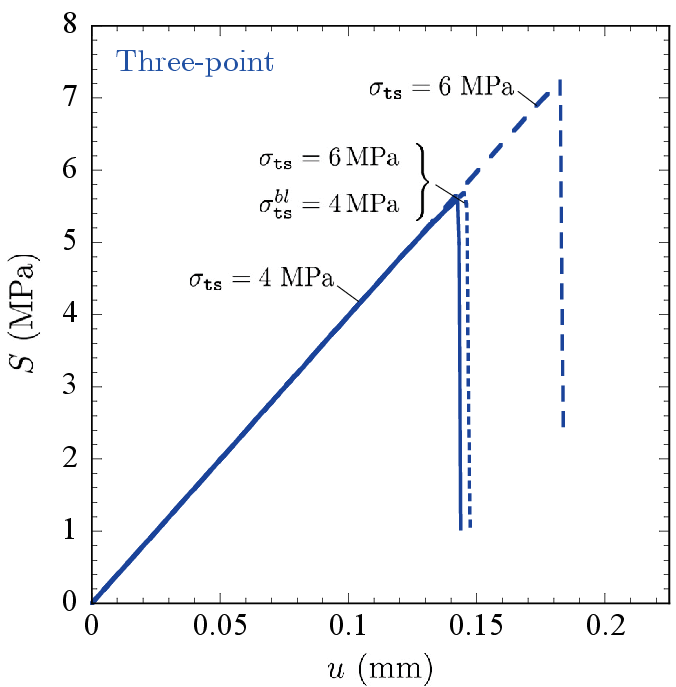}
\caption{{\small The global response ($S$ vs. $u$) of the beam under three-point bending for three mortars: two with different uniform uniaxial tensile strengths, $\sts=4$ MPa and $\sts=6$ MPa, and one with bulk uniaxial tensile strength $\sts=6$ MPa and a $25$-mm-thick boundary layer with uniaxial tensile strength $\sts^{bl}=4$ MPa. For all three mortars, the remaining constants are the same as in Fig.~\ref{Fig7}: $E=27$ GPa, $\nu=0.20$, $\scs=60$ MPa, and $G_c=30$ N/m.}}\label{Fig17}
\end{figure}
%

Figures \ref{Fig16} and \ref{Fig17} present comparisons for the global responses of the beam under four-point and three-point bending for three material configurations: the original mortar with uniform strength ($\sts=4$ MPa) previously shown in Fig.~\ref{Fig7}, a mortar with a higher uniform strength ($\sts=6$ MPa), and a mortar featuring a 25-mm-thick boundary layer. In this final case, the bulk uniaxial tensile strength is set to $\sts=6$ MPa, while the uniaxial tensile strength in the boundary layer is maintained at the original bulk value of $\sts^{bl}=4$ MPa; see Fig.~\ref{Fig15}.

A quick glance at the results suffices to recognize that, for the representative cases considered here, the fracture nucleation is primarily governed by the uniaxial tensile strength $\sts^{bl}$ in the boundary layer, with the bulk uniaxial tensile strength $\sts$ playing a secondary role. This finding is consistent with the fact that fracture nucleation occurs at the boundary and is immediately followed by a very rapid propagation for both bending tests. Importantly, this implies that experimentally measured maximum global stresses, $S^{4p}_{max}$ and $S^{3p}_{max}$, contain information primarily regarding $\sts^{bl}$ rather than $\sts$. In cases where $\sts^{bl}$ deviates significantly from $\sts$, alternative testing methods would thus be necessary to accurately characterize the bulk uniaxial tensile strength.

\subsubsection{Spatial stochasticity of the strength in the boundary layer}

Based on the preceding analysis, we now investigate how stochastic spatial variations in the uniaxial tensile strength $\sts^{bl}$ in the boundary layer influence the response of the beam. For simplicity, we assume random fluctuations between three discrete values:
\begin{equation*}
\sts^{bl}=\{3.6, 4.0, 4.4\}\, {\rm MPa}.
\end{equation*}
These variations are applied across equiaxed regions, each with a characteristic size of $6.25\varepsilon=25$ mm, distributed throughout the boundary layer. The remaining material constants are the same as in Subsection \ref{SubS-4p-3p}: $E=27$ GPa, $\nu=0.20$, $\sts=4$ MPa, $\scs=60$ MPa, and $G_c=30$ N/m.

%
\begin{figure}[b!]
\centering
\includegraphics[width=0.68\linewidth]{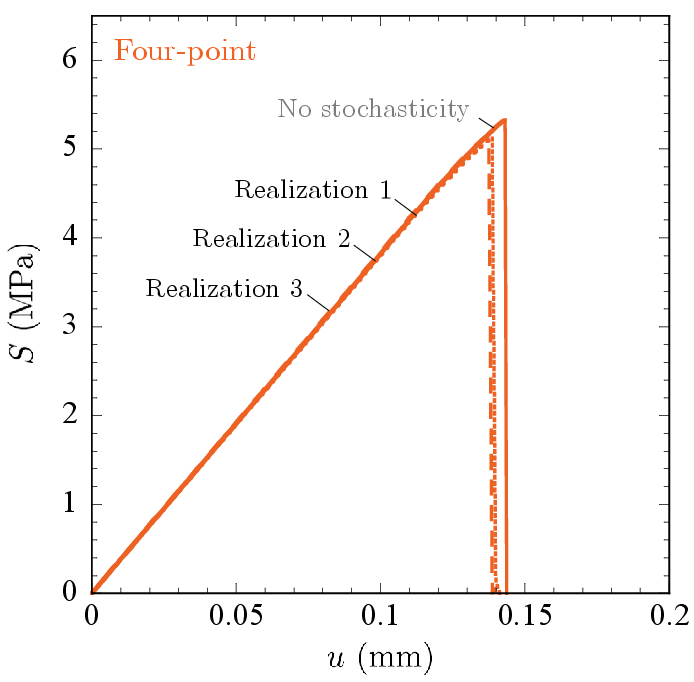}
\caption{{\small The global response ($S$ vs. $u$) of the beam under four-point bending for mortars with three different realizations of strength stochasticity. The response for mortar without stochasticity (with spatially uniform strength $\sts=4$ MPa throughout) is also included for direct comparison.}}\label{Fig18}
\end{figure}
%

%
\begin{figure}[t!]
\centering
\includegraphics[width=0.68\linewidth]{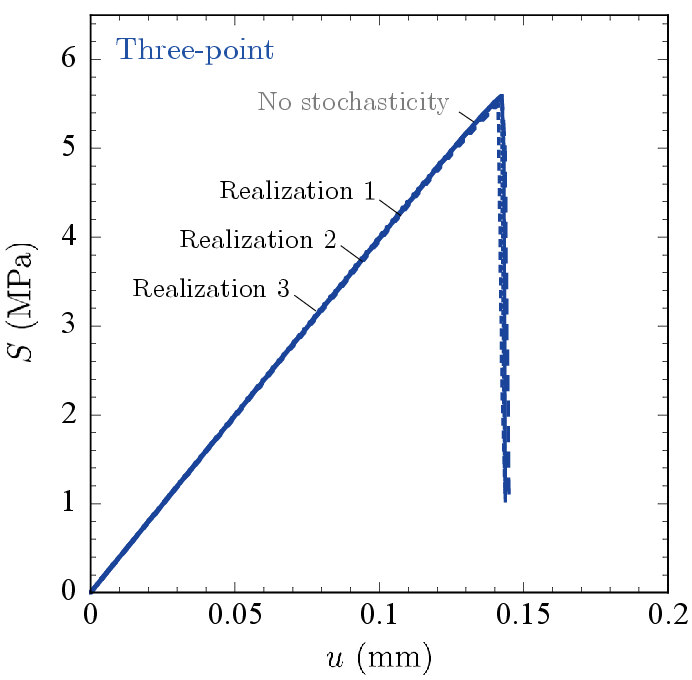}
\caption{{\small The global response ($S$ vs. $u$) of the beam under three-point bending for mortars with three different realizations of strength stochasticity. The response for mortar without stochasticity (with spatially uniform strength $\sts=4$ MPa throughout) is also included for direct comparison.}}\label{Fig19}
\end{figure}
%

Figures \ref{Fig18} and \ref{Fig19} present results for the global responses under four-point and three-point bending of the beam with three different realizations of strength stochasticity. For direct comparison, the corresponding response of the beam made of a mortar with uniform strength is also included in both figures.

The primary observation from these results is that under four-point bending fracture nucleation is governed by the minimum value $\sts^{bl}=3.6$ MPa of the uniaxial tensile strength within the boundary layer, while under three-point bending it is governed by the average value $\sts^{bl}=4$ MPa. The former results also serve to reveal that a reduction in the minimum value of $\sts^{bl}$ leads to a non-proportional reduction in the value of the maximum global stress $S_{max}$ at which fracture nucleates under four-point bending. In particular, in this case, a 10\% reduction in $\sts^{bl}$ (from 4 MPa to 3.6 MPa) leads to a reduction in $S_{max}$ of approximately 4\% (from 5.32 MPa to, on average, 5.12 MPa). 

Finally, we note that strength stochasticity significantly influences the specific location of fracture nucleation under four-point bending, whereas the nucleation site under three-point bending remains largely unaffected by such randomness. Across all three realizations for three-point bending, the crack consistently nucleated at around the beam's center. On the other hand, under four-point bending, the crack nucleated at varying off-center positions between the inner supports depending on the realization. Specifically, for the three cases studied here, nucleation occurred at $X_1=80, 145$, and $140$ mm.

\section{Summary and simple formulas to deduce the uniaxial tensile strength from bending tests}\label{Sec: Final Comments}

In this paper, we have presented a comprehensive 3D quantitative analysis of when and where fracture nucleates and propagates in a beam under four-point and three-point bending. The results have brought resolution to two long-standing problems of fundamental and practical significance: ($i$) how to properly interpret peak load measurements, most often  reported as the flexural strength, and ($ii$) why four-point and three-point bending generally yield different results for the flexural strength. 

Specifically, the main findings can be summarized as follows:
\begin{enumerate}[label=\textbf{\roman*.}]
\item{The maximum global stress $S_{max}$, or flexural strength, at which fracture nucleates in a beam under either type of bending can be effectively expressed in the form
\begin{equation}\label{Smax-Final}
S_{max}=g\left(\dfrac{L_s}{H},\dfrac{D}{H},\dfrac{L_l}{H},\dfrac{E G_c}{H\sts^2}\right)\sts,
\end{equation}
where $\sts$ is the uniaxial tensile strength within the boundary layer\footnote{Throughout this final section, we drop the use of the boundary-layer superscript ``$bl$'' for notational simplicity.} and $g$ is a dimensionless function of: 
\begin{itemize}
\item{the geometry of the beam via the span-to-height ratio $L_s/H$ and the depth-to-height ratio $D/H$;}
\item{the loading configuration via the loading-span-to-height ratio $L_l/H$; and}
\item{the beam material properties via the material length scale $E G_c/\sts^2$, normalized by the height $H$ of the beam.}
\end{itemize}
Notably, $S_{max}$ is effectively independent of the Poisson's ratio $\nu$, the uniaxial tensile strength in the bulk (away from the boundary layer), and the topology of the strength surface $\mathcal{F}(\boldsymbol{\sigma})=0$ outside uniaxial tension. 
}

\item{The dependence of $S_{max}$ on the material length scale $E G_c/\sts^2$ is a telltale sign of the process of fracture nucleation. As the load increases, fracture nucleation occurs when the strength surface $\mathcal{F}(\boldsymbol{\sigma})=0$ of the material is exceeded over a region of sufficient height from the boundary such that forming a crack within that region is energetically favorable. This critical height is a material-specific length that scales with $EG_c/\sts^2$.
}

\item{The dependence of $S_{max}$ on the loading-span-to-height ratio $L_l/H$ reflects how different loading configurations generate different stress fields. These fields lead to different violations of the strength surface, which in turn lead to different maximum global stresses at which fracture nucleates. For the two loading configurations of interest here, the function $g$ in expression (\ref{Smax-Final}) for four-point bending ($L_l/H=2$) happens to bound from below the three-point bending function ($L_l/H=0$), this for fixed $L_s/H$ and $D/H$ across all values of  $EG_c/(H\sts^2)$. Consequently, fracture occurs at lower $S_{max}$ under four-point bending than under three-point bending.
}

\item{In expression (\ref{Smax-Final}), the function $g\geq 1$ is a monotonically increasing function of its last argument, the normalized material length scale $EG_c/(H\sts^2)$. This implies that:
\begin{itemize}
\item{$S_{max}\geq\sts$, irrespective of the beam geometry, loading configuration, and material; and}
\item{for fixed span-to-height ratio ($L_s/H$), fixed depth-to-height ratio ($D/H$), and fixed loading-span-to-height ratio ($L_l/H$), fracture occurs at lower $S_{max}$ in beams of larger height $H$.}
\end{itemize}

}

\item{In practice, any given beam exhibits stochastic spatial variation in its uniaxial tensile strength. The value of $\sts$ in expression (\ref{Smax-Final}) should be viewed differently depending on the loading configuration. For four-point bending ($L_l/H=2$), it should be considered as the minimum strength within the boundary layer where fracture nucleates. For three-point bending ($L_l/H=0$), on the other hand, it should be considered as the average strength within the boundary layer.

}

\end{enumerate}

\subsection{Formulas}

We close by proposing simple explicit formulas for the function $g$ in expression (\ref{Smax-Final}) that can be readily utilized by practitioners to deduce the uniaxial tensile strength $\sts$ (within the boundary layer) directly from measurements of the flexural strength $S_{max}$ in four-point and three-point bending tests.\footnote{Of course, being dependent not just on $\sts$ but also on $E$ and $G_c$, the proposed formulas could be alternatively utilized to deduce the critical energy release rate $G_c$ or Young's modulus $E$ of the material.}

The proposed formulas were derived by fitting predictions generated by the phase-field theory across a range of beam sizes and material constants. The scope of this study was restricted to beams with a span-to-height ratio of $L_s/H=4$ and a depth-to-height ratio of $D/H=1$. Consequently, the resulting expressions are applicable to geometries maintaining these proportions. We conducted ten simulations for each loading configuration. The results for four-point bending ($L_l/H=2$) and three-point bending ($L_l/H=0$) are summarized in Table \ref{Table4} and Table \ref{Table5}, respectively. Both datasets are well approximated by a power-law relationship. Specifically, for four-point bending, the relationship is given by:
\begin{equation*}
S^{4p}_{max}=\left(1+\alpha\left(\dfrac{E G_c}{H\sts^2}\right)^\beta\right)^{\gamma}\sts \quad{\rm with}\quad\left\{\begin{array}{l}
\hspace{-0.15cm}\alpha=1.567\vspace{0.2cm}\\
\hspace{-0.15cm}\beta=0.775\vspace{0.2cm}\\
\hspace{-0.15cm}\gamma=0.678\end{array}\right. .
\end{equation*}
Similarly, for three-point bending:
\begin{equation*}
S^{3p}_{max}=\left(1+\alpha\left(\dfrac{E G_c}{H\sts^2}\right)^\beta\right)^{\gamma}\sts \quad{\rm with}\quad\left\{\begin{array}{l}
\hspace{-0.15cm}\alpha=0.965\vspace{0.2cm}\\
\hspace{-0.15cm}\beta=0.614\vspace{0.2cm}\\
\hspace{-0.15cm}\gamma=1\end{array}\right. .
\end{equation*}

\begin{table}[t!]\centering\small
\caption{Values predicted by the phase-field theory for the maximum global stress $S^{4p}_{max}$ at which fracture nucleates in beams, of various dimensions ($L_s/H=4$, $D/H=1$) and various material properties ($E=27$ GPa, $\nu=0.20$, $\scs=60$ MPa), under four-point bending ($L_l/H=2$).}
\begin{tabular}{cccc?cc}
\toprule
$H$                     & $\sts$                    & $G_c$                     &  $E G_c/(H\sts^2)$  & $S^{4p}_{max}$       &  $S^{4p}_{max}/\sts$      \\
\footnotesize{(m)}      & \footnotesize{(MPa)}      & \footnotesize{(N/m)}      &                     & \footnotesize{(MPa)} &   \\
\midrule
\midrule
0.2          & 4            & 5        & 0.042                      & 4.296   &    1.074       \\
\midrule
\midrule
0.2          & 4            & 15        & 0.127                       & 4.812   &    1.203        \\
\midrule
0.2        & 5.657        & 30        & 0.127                        & 6.794    &    1.201       \\
\midrule
0.4          & 4           & 30        & 0.127                        & 4.780   &    1.195       \\
\midrule
\midrule
0.2          & 4            & 30        & 0.253                       & 5.324    &   1.331        \\
\midrule
0.1         & 8            & 60        & 0.253                       & 10.624    &    1.328       \\
\midrule
0.2          & 1.633          & 5        & 0.253                      & 2.187    &   1.339        \\
\midrule
\midrule
0.2        & 4           & 60        & 0.506                       & 6.288     &   1.572      \\
\midrule
0.1        & 2.828         & 15        & 0.506                       & 4.448   &  1.573         \\
\midrule
0.2        & 2.828            & 30        & 0.506                       & 4.491  &  1.588           \\
\bottomrule
\end{tabular} \label{Table4}
\end{table}
\begin{table}[t!]\centering\small
\caption{Values predicted by the phase-field theory for the maximum global stress $S^{3p}_{max}$ at which fracture nucleates in beams, of various dimensions ($L_s/H=4$, $D/H=1$) and various material properties ($E=27$ GPa, $\nu=0.20$, $\scs=60$ MPa), under three-point bending ($L_l/H=0$).}
\begin{tabular}{cccc?cc}
\toprule
$H$                     & $\sts$                    & $G_c$                     &  $E G_c/(H\sts^2)$  & $S^{3p}_{max}$       &  $S^{3p}_{max}/\sts$      \\
\footnotesize{(m)}      & \footnotesize{(MPa)}      & \footnotesize{(N/m)}      &                     & \footnotesize{(MPa)} &   \\
\midrule
\midrule
0.2          & 4            & 5        & 0.042                      & 4.628   &    1.157        \\
\midrule
\midrule
0.2          & 4            & 15        & 0.127                       & 5.104   &  1.276          \\
\midrule
0.2        & 5.657        & 30        & 0.127                        & 7.230    &  1.278         \\
\midrule
0.4          & 4           & 30        & 0.127                        & 5.124   &  1.281          \\
\midrule
\midrule
0.2          & 4            & 30        & 0.253                       & 5.592    &    1.398          \\
\midrule
0.1         & 8            & 60        & 0.253                       & 11.240    &  1.405         \\
\midrule
0.2          & 1.633          & 5        & 0.253                      & 2.280   &  1.396         \\
\midrule
\midrule
0.2        & 4           & 60        & 0.506                       & 6.552    &  1.638       \\
\midrule
0.1        & 2.828         & 15        & 0.506                       & 4.658    &  1.647         \\
\midrule
0.2        & 2.828            & 30        & 0.506                       & 4.638  &  1.640          \\
\bottomrule
\end{tabular} \label{Table5}
\end{table}

Interestingly, the above formulas are similar in functional form to various empirical relations that have been proposed in the literature over the years to fit experimental data; see, e.g., \citep{Bazant01}.

We conjecture that this type of power-law relation remains valid for arbitrary beam geometries and loading configurations. Specifically, variations in the span-to-height ratio ($L_s/H$), depth-to-height ratio ($D/H$), and loading-span-to-height ratio ($L_l/H$) are expected to influence the specific values of the coefficient $\alpha$ and the exponents $\beta$ and $\gamma$ without altering the fundamental functional form. Characterizing $\alpha$, $\beta$, and $\gamma$ as explicit functions of these geometric parameters is left as a future exercise.

\section*{Acknowledgements}

\noindent This work was supported by the National Science Foundation through the Grant DMS--2308169. This support is gratefully acknowledged.


\end{document}